\documentclass[twocolumn]{aastex7}
\usepackage{amsmath}
\usepackage{multirow}
\usepackage{algorithm2e}
\usepackage{enumitem}
\usepackage{gensymb}

\begin{document}

\title{Spectuner-D1: Spectral Line Fitting of Interstellar Molecules Using Deep Reinforcement Learning}

\author[0000-0002-7716-1094]{Yisheng Qiu}
\affiliation{Research Center for Astronomical computing, Zhejiang Laboratory, Hangzhou 311121, China}
\email{yishengq@zhejianglab.org}

\author[0000-0002-1466-3484]{Tianwei Zhang}
\affiliation{Research Center for Astronomical computing, Zhejiang Laboratory, Hangzhou 311121, China}
\email{twzhang@zhejianglab.org}

\author[0000-0002-5286-2564]{Tie Liu}
\affiliation{Shanghai Astronomical Observatory, Chinese Academy of Sciences, 80
Nandan Road, Shanghai 200030, China}
\email{liutie@shao.ac.cn}

\author{Fengyao Zhu}
\affiliation{Research Center for Astronomical computing, Zhejiang Laboratory, Hangzhou 311121, China}
\email{zhufy@zhejianglab.org}

\author[0009-0000-5764-8527]{Dezhao Meng}\affiliation{Xinjiang Astronomical Observatory, Chinese Academy of Sciences, Urumqi 830011, People’s Republic of China}
\affiliation{University of Chinese Academy of Sciences, Beijing 100080,
People’s Republic of China}
\affiliation{Shanghai Astronomical Observatory, Chinese Academy of Sciences, 80
Nandan Road, Shanghai 200030, China}
\email{mengdezhao@xao.ac.cn}

\author[0009-0000-6108-2730]{Huaxi Chen}
\affiliation{Research Center for Astronomical computing, Zhejiang Laboratory, Hangzhou 311121, China}
\email{chenhuaxi@zhejianglab.org}

\author[0000-0002-9277-8025]{Thomas M{\"o}ller}
\affiliation{I. Physikalisches Institut, Universit{\"a}t zu K{\"o}ln, Z{\"u}lpicher Stra{\ss}e 77, 50937 Cologne, Germany}
\email{moeller@ph1.uni-koeln.de}

\author[0000-0003-2141-5689]{Peter Schilke}
\affiliation{I. Physikalisches Institut, Universit{\"a}t zu K{\"o}ln, Z{\"u}lpicher Stra{\ss}e 77, 50937 Cologne, Germany}
\email{schilke@ph1.uni-koeln.de}

\author[0000-0003-4811-2581]{Donghui Quan}
\affiliation{Research Center for Astronomical computing, Zhejiang Laboratory, Hangzhou 311121, China}
\email{donghui.quan@zhejianglab.org}

 
\begin{abstract}
Spectral lines from interstellar molecules provide crucial insights into the physical and chemical conditions of the interstellar medium. Traditional spectral line analysis relies heavily on manual intervention, which becomes impractical when handling the massive datasets produced by modern facilities like ALMA. To address this challenge, we introduce a novel deep reinforcement learning framework to automate spectral line fitting. Using observational data from ALMA, we train a neural network that maps both molecular spectroscopic data and observed spectra to physical parameters such as excitation temperature and column density. The neural network predictions can serve as initial estimates and be further refined using a local optimizer. Our method achieves consistent fitting results compared to global optimization with multiple runs, while reducing the number of forward modeling runs by an order of magnitude. We apply our method to pixel-level fitting for an observation of the G327.3-0.6 hot core and validate our results using \textsc{xclass}. We perform the fitting for typical complex organic molecules of hot cores, including CH$_3$OH, CH$_3$OCHO, CH$_3$OCH$_3$, C$_2$H$_5$CN, and C$_2$H$_3$CN. For a 100 $\times$ 100 region covering 5 GHz bandwidth, the fitting process requires 4.9 to 41.9 minutes using a desktop with 16 cores and one consumer-grade GPU card.
\end{abstract}



\section{Introduction} 
Spectral lines from interstellar species provide a crucial probe of the physical and chemical conditions in the interstellar medium (ISM). The analysis of spectral lines involves several key steps: querying transitions from spectroscopic databases, matching these transitions to observed spectral lines, and determining a set of physical parameters that best reproduce the observed spectrum \cite[e.g.][]{1999ApJ...517..209G}. Traditionally, spectral line analysis has been a labor-intensive process, requiring significant expertise and time to manually match observed features with known transitions and iteratively adjust parameters to achieve satisfactory fits. Although several analysis tools have been developed over the past decade, e.g. \textsc{weeds} \citep{2011A&A...526A..47M}, \textsc{cassis} \citep{2015sf2a.conf..313V}, \textsc{xclass} \citep{2017A&A...598A...7M}, \textsc{madcuba} \citep{2019A&A...631A.159M}, and \textsc{pyspeckit} \citep{2022AJ....163..291G}, manual intervention remains common. As the volume and complexity of observational data continue to increase due to advanced observational facilities like the Atacama Large Millimeter/submillimeter Array (ALMA) \citep{2020arXiv200111076C}, there is a growing need for automated and efficient methods to perform this analysis, especially in line-rich regions such as hot cores and hot corinos \citep[see][for a review]{2020ARA&A..58..727J}.
\par
Recent advances in machine learning offer a promising opportunity to transform spectral analysis workflows. For instance, machine learning algorithms have been used to establish the relation between spectral line properties to the underlying physical parameters \citep{2024A&A...691A.109E,2025A&A...702A..71G,2025A&A...698A.286M}. \citet{2025arXiv251009119K} trained a convolutional neural network for identification of complex organic molecules (COMs) using their spectral line features. In addition, machine learning techniques were applied to explore the chemical inventories of molecular clouds \citep{2021ApJ...917L...6L,2025ApJ...985..123T} and to investigate the interstellar chemical reaction networks \citep{2022A&A...666A..45V,2023MNRAS.526..404H}.
\par
In this paper, we present a deep learning framework to automate and enhance the spectral line fitting process. Specifically, we aim to train a neural network that can infer the physical parameters of molecule emission such as excitation temperature and column density from the observed spectrum and the molecular transition properties within the observed spectral windows. We choose deep learning over traditional machine learning methods such as random forests \citep{breiman2001random} because it scales more effectively with large datasets and is better suited to handling the massive data volumes produced by ALMA and other modern observatories. The inferred physical parameters of the neural network are based on a one-dimensional radiative transfer model that assumes local thermodynamic equilibrium (LTE). While this assumption may introduce bias \citep[e.g.][]{2024A&A...686A.255R}, the simplicity of the LTE model enables efficient computation of model spectra, which significantly reduces the time required for training and testing and makes the project feasible.
\par
This work employs deep reinforcement learning (DRL) \citep[e.g.][]{mnih2015human,schulman2017proximal} to train the neural network. To improve the generalization of the neural network, we seek to train it using real ALMA data. However, vast amounts of labeled data are unavailable for this task. This challenge could be addressed using DRL. Unlike supervised learning, DRL does not require labeled data and instead discovers optimal results during training. In our approach, training proceeds by interacting with a reward model and receiving feedback through reward signals that reflect fit quality. The reward model uses the physical parameters inferred by the neural network to compute the model spectra under LTE conditions and estimates the fit quality by comparing them to the observed spectrum.
\par
We propose that our neural network can be combined with traditional fitting methods. Specifically, the neural network generates an initial parameter estimate, which is then refined by a local optimizer to produce high-precision fits. This hybrid approach could significantly reduce manual intervention in spectral line analysis while preserving the accuracy of the parameter estimates.
\par
Our neural network-based approach can be efficiently applied to pixel-by-pixel fitting of spectral line cubes, which is a challenging task. When using local optimization algorithms, providing good initial parameter estimates for each pixel is non-trivial, and poor initial guesses can significantly affect the the fitting results. For instance, to improve the fitting, \cite{2024ApJ...965...14E} proposed using the results from previously fitted neighboring pixels to initialize the parameters for subsequent fits. On the other hand, while global optimization methods could lead to more robust fitting results, they are time consuming and require large number of forward modeling runs. In contrast, our neural network can provide good initial guesses for all pixels, leading to efficient and robust pixel-by-pixel fitting.
\par
This paper is organized as follows. Section \ref{sec:method} introduces our methodology, including the description of the training data, along with the training and evaluation methods. Section \ref{sec:results} presents the training results. In Section \ref{sec:app}, we apply our method to pixel-level spectral fitting. We then discuss our methodology in Section \ref{sec:discuss}. Finally, this work is summarized in Section \ref{sec:summary}.
\par
The code for the neural network-based fitting method is incorporated into our \textsc{spectuner} Python package, which is publicly available on Github\footnote{\url{https://github.com/yqiuu/spectuner}}. The neural network weights is publicly available on Hugging face\footnote{\url{https://huggingface.co/yqiuu/Spectuner-D1}}.

\begin{figure*}
	\includegraphics[width=\textwidth]{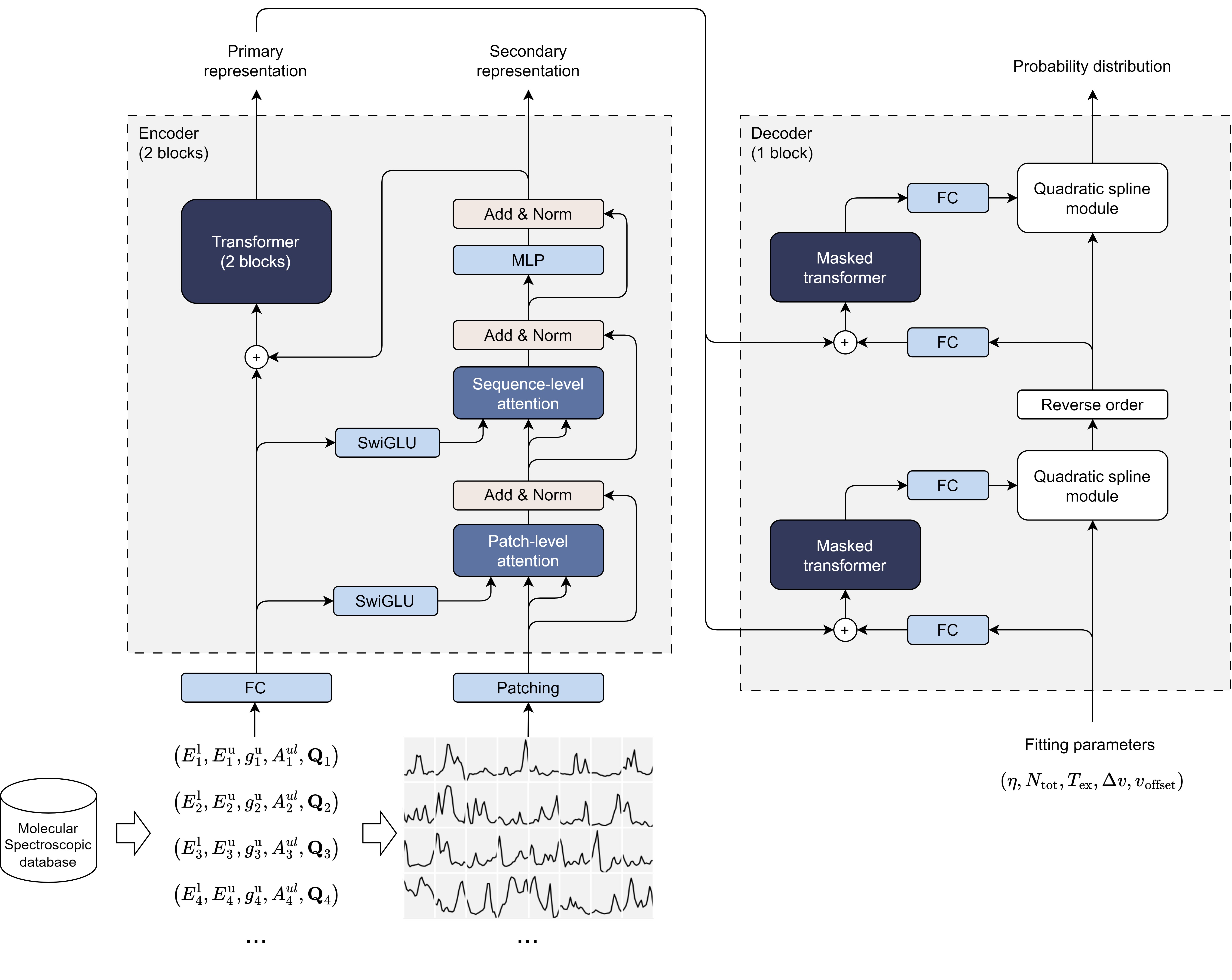}
	\centering
    \caption{Architecture of the proposed neural network introduced in Section \ref{sec:architecture}. Here, FC stands for the fully connected layer, MLP stands for the multi-layer perceptron, and SwiGLU stands for the SwiGLU activation function \citep{2020arXiv200205202S}.}
    \label{fig:architecture}
\end{figure*}

\section{Methodology} \label{sec:method}
\subsection{Network inputs} \label{sec:input}
This work aims to train a neural network for spectral line fitting, and therefore the network should include all the information typically used in such fitting procedures. For observational data, the format is significantly different from that used in traditional deep learning applications such as computer vision (CV) and natural language processing (NLP). An observed spectrum consists of multiple spectral windows. The number of spectral windows, the frequency range of each window, and the number of frequency channels in each window can all vary between observations. We also need to include information about the molecular species being fitted. To overcome the challenges above, this work proposes a method that converts the input data into a pair of length-variable sequences, which can be efficiently processed using neural networks.
\par
Given an observed spectrum with multiple spectral windows and a molecule identifier, we first query the molecular spectroscopic database to find all transitions of the molecule within those frequency ranges, resulting in a $N_\text{trans} \times F_\text{trans}$ array, denoted as $\textbf{d}_\text{trans}$, where $N_\text{trans}$ is the number of transitions and $F_\text{trans}$ is the number of features of each transition. For example, if the input spectral windows span 97 - 99 GHz and 100 - 102 GHz, and the target molecule is CH$_3$OH, v=0, we query the molecular spectroscopic database for all transitions of CH$_3$OH, v=0 within these frequency ranges, and may find 25 transitions; in this case $N_\text{trans}=25$. Data are grouped into batches during training. A single batch may contain multiple species, e.g. CO, v=0, CH$_3$OH, v=0, and H$_2$CO, v=0, and the corresponding spectral windows may also be different. Therefore, the number of transitions $N_\text{trans}$ varies across these input data. This situation is common in NLP, where the number of transitions is analogous to sentence length, and can be effectively handled using transformer-based neural networks \citep{NIPS2017_3f5ee243}. We adopt this network architecture in our work, as described in Section \ref{sec:architecture}.
\par
We include the following features for each transition: the energy of the lower state $E_\text{l}$, the energy of the upper state $E_\text{u}$, the upper state degeneracy $g_\text{u}$, the Einstein coefficient $A_\text{ul}$, and the partition function $\textbf{Q}$ at different temperatures. The molecular spectroscopic data are from the Virtual Atomic and Molecular Data Center (VAMDC) \citep{2016JMoSp.327...95E}, which contains entries from the Cologne Database for Molecular Spectroscopy (CDMS) \citep{2001A&A...370L..49M,2005JMoSt.742..215M} and the Jet Propulsion Laboratory (JPL) database \citep{1998JQSRT..60..883P}.
\par
Normalizing input features is a standard practice in neural network training. This work applies Z-score normalization to the transition features, which transforms the data to have zero mean and unit variance. For $A_\text{ul}$ and $g_\text{u}$, normalization is conducted in logarithmic space, whereas for $E_\text{l}$, $E_\text{u}$, and $\textbf{Q}$, normalization is applied after transformation using the inverse hyperbolic sine (arcsinh) function. We have found exact zero values for $E_\text{l}$, $E_\text{u}$, and $\textbf{Q}$ in the spectroscopic database, and therefore suggest using the arcsinh function for better numerical stability. We expect that it performs similarly to the logarithmic transform. Normalization using the arcsinh function depends on the unit of the input values. The units of $E_\text{l}$ and $E_\text{u}$ are converted to Kelvin before applying the arcsinh function, while the partition function values are dimensionless.
\par
The second input to the neural network is derived from the observed spectrum using the transition frequency. For each transition, we extract frequency channels within a window centered on the transition frequency and spanning a velocity width of 200 km/s. The extracted data include the following features:
\begin{enumerate}[itemsep=0pt]
    \item Frequency.
    \item Intensity.
    \item An indicator (0 or 1) of whether the channel has a peak. 
\end{enumerate}
The third feature is derived using the approach described in \citet{2025ApJS..277...21Q}. Specifically, we use the \textit{find\_peaks} function implemented by \textit{scipy}, which identifies all local maxima in a 1D input array based on user-specified criteria. The most relevant criterion of the algorithm is the prominence threshold. Prominence is a property of a peak, defined as the vertical distance from the peak height to the lowest contour line that does not enclose any higher peak. The algorithm ignores any peak with a prominence below the specified threshold. In this work, we set the prominence threshold to four times the root mean square (RMS) noise of the input spectrum. The method to estimate the noise levels is described in Section \ref{sec:preprocessing}. In addition, the peak finding algorithm assumes that the input spectrum has a flat baseline. We have checked that this assumption is valid for the vast majority of the observational data used in this study.
\par
The frequency resolution of the observed spectra differs among spectral windows and observations, and therefore the extracted data for each transition have different lengths. We require the data to have the same length for use in our neural network. Accordingly, we map the extracted data onto a fixed-length grid, producing a $N_\text{trans} \times N_\text{grid} \times F_\text{spec}$ array, denoted as $\textbf{d}_\text{spec}$, where $N_\text{grid}$ is the number of grid points and $F_\text{spec}$ is the number of features defined for the spectral data. The grid is defined by velocity shifts relative to the transition frequency. We use linear interpolation for intensities and the nearest-neighbor method for peak flags. Grid points outside the original spectral window are assigned zero intensity. We also include an indicator (0 or 1) specifying whether each grid point lies outside the original spectral window. Furthermore, the following properties, constant across all grid points, are included:
\begin{enumerate}[itemsep=0pt]
    \item Ratio of the grid frequency resolution to that of the original spectrum.
    \item Continuum temperature.
    \item RMS noise.
    \item Beam size.
\end{enumerate}
As a standard practice, we normalize the numerical features before training. Velocity shifts are divided by 200 km/s, i.e. the window width. The arcsinh transformation is applied to the intensity of the observed spectrum, continuum temperature, and RMS noise, all expressed in Kelvin. For the beam size, we apply min-max normalization in logarithmic space, with minimum and maximum values of 0.01$''$ and 10$''$. While other normalization choices are possible, we found the above approach effective in practice.
\par
In addition, we set $N_\text{grid} = 1920$, corresponding to frequency resolutions of 80 kHz at 1.3 mm and 35 kHz at 3 mm. This value is chosen so that the frequency resolutions of the grid is finer than all observational data used in this work. However, larger $N_\text{grid}$ is not explored due to computational speed and GPU memory limitations.

\begin{table}
\caption{Summary of the fitting parameters in our spectral line model described in Section \ref{sec:sl_model}.}
\label{tab:params}
\centering
\begin{tabular}{cccc}
\hline
\hline
    Name & Unit & Scale & Bound \\
\hline
    Filling factor $\eta$ & $''$ & log & -3 - -0.00043 \\
    Excitation temperature $T_\text{ex}$ & K & linear & 1 - 1000 \\
    Column density $N_\text{tot}$ & cm$^{-2}$ & log & 12 - 22 \\
    Velocity width $\Delta v$ & km/s & log & -0.5 - 1.5 \\
    Velocity offset $v_\text{LSR}$ & km/s & linear & -12 - 12 \\
\hline
\end{tabular}
\end{table}

\subsection{Network architecture} \label{sec:architecture}
Our network architecture, which consists of an encoder and a decoder, is illustrated in Figure \ref{fig:architecture}. The neural network only uses basic modules in deep learning, including the fully connected (FC) layer, the multilayer perceptron (MLP), the patching layer in the vision transformer (ViT) \citep{dosovitskiy2021an}, the SwiGLU activation function \citep{2020arXiv200205202S}, the multi-head attention module \citep{NIPS2017_3f5ee243} and the transformer layer \citep{NIPS2017_3f5ee243}. Readers may also refer to \cite{Goodfellow-et-al-2016} for the relevant concepts of deep learning.
\par
The encoder of our neural network converts the input variable-length sequences of molecular spectroscopic and observational data into dense vectors. This process is analogous to models in natural language processing (NLP) that learn vector embeddings of sentences and paragraphs, such as BERT \citep{devlin-etal-2019-bert}. A transformer-based architecture is particularly suitable for this purpose, and the so-called [CLS] token \citep{devlin-etal-2019-bert} can be used to represent aggregate sequence-level information. In astronomy, transformer-based neural networks have been applied to learning the representations of galaxy images and spectra \citep[e.g.][]{2024MNRAS.531.4990P,2025AJ....170...28R}.
\par
For the sequence of the observed spectrum, we first split the sub-window spectrum into several patches using the patching layer in ViT \citep{dosovitskiy2021an}. The same patching operation is applied across the sequence dimension. We then extract the features of the observed spectrum using a 3D transformer, with the query vector generated based on the representation of the molecular transition data. For the patch-level attention, the query vector is generated using only the embedding of the [CLS] token. In other words, the same query vector is shared across the sequence dimension. For the sequence-level attention, the query vector is generated per token but shared across the patch dimension. For the sequence of the molecular transition data, we employ two transformer layers \citep{NIPS2017_3f5ee243} to model the correlation between the transitions. Unlike in NLP, our sequences are unordered, and therefore no positional embeddings are applied to the transformers. Finally, the [CLS] token embedding corresponding to the molecular spectroscopic data is passed to the decoder.
\par
The decoder of the neural network is a normalizing flow \citep[e.g.][]{NIPS2017_6c1da886,papamakarios2021normalizing}, which models conditional probability distributions by iteratively applying reversible transformations to a simple multi-variate distribution. Normalizing flows allow both density estimation and sampling from the modeled distribution. This capability is essential for calculating the policy gradient during our training. In astronomy, normalizing flows have been applied to parameter estimation in cosmology \citep{2019MNRAS.488.4440A} and gravitational wave signal analysis \citep{2023PhRvD.108d2004B}. 
\par
This work employs the autoregressive flow, which is composed of a parameterized transformation and a conditioner \citep{papamakarios2021normalizing}. We adopt the piecewise quadratic transformation proposed by \cite{mueller18neural-v2} and the masked transformer \citep{NIPS2017_3f5ee243} for the conditioner. Similar neural networks were presented by \cite{patacchiola2024transformer}, who showed that the use of transformers could lead to better results in their benchmarks. When using the transformer, each input parameter is treated as a one-dimensional token, which is projected using a fully connected (FC) layer to match the dimension of the encoder embedding. Then, the embedding of each parameter is combined with the same encoder embedding via addition. For each token, the piecewise quadratic transformation is parameterized using $2N_\text{bin} + 1$ dimensions, where $N_\text{bin}$ is the number of bins used to partition the domain of the piecewise polynomials. Consequently, we employ a FC layer to project the output of the masked transformer to the required $2N_\text{bin} + 1$ dimensions. In this work, the decoder predicts five parameters, which are explained in Section \ref{sec:sl_model}. The output of the quadratic spline module is bounded, and the adopted bounds are given in Table \ref{tab:params}.
\par
After several experiments, we suggest the following hyperparameters for our neural network. For the multi-head attention and transformer layers,  we adopt 768 embedding dimensions and 8 heads. The hidden size in the MLP is set to be $768 \times 8/3 = 2048$, a common practice when using the SwiGLU activation function \citep{2020arXiv200205202S}. For the sequence of the observed spectrum, the sub-windows are split into 8 patches. We adopt 32 bins for the quadratic spline module. In addition, as illustrated in Figure \ref{fig:architecture}, we use 2 blocks and 1 block of the corresponding aggregated modules for the encoder and decoder, respectively. Our neural network has 89 million parameters in this configuration.

\subsection{The spectral line fitting module}
The spectral line fitting module takes the fitting parameters as input, calculates the model spectrum, compares it with the observed spectrum, and finally produces a similarity score. The similarity score is used to compute the reward for network training.

\subsubsection{The spectral line model} \label{sec:sl_model}
This work adopts a one-dimensional LTE spectral line model for a molecule or isotopologue in a single vibrational state with a single velocity component \citep[see][for a nice derivation]{2017A&A...598A...7M}. The model is characterized by five fitting parameters, the same quantities that our neural network predicts. Although real observed spectra typically contain emission from multiple species and vibrational states, fitting results for an individual species in a single vibrational state provide a valuable starting point for more detailed analysis. The resulting parameters could, for example, be used as initial values in subsequent joint fitting procedures.
\par
The adopted spectral line model is computed by \citep{2017A&A...598A...7M}

{
\allowdisplaybreaks
\begin{align}
    J_\nu &= \eta(\theta) \left(S_\nu  - J^{bg}_\nu 
    \right) \left(1 - e^{-\tau_\nu}\right), \label{eqn:slm_first} \\
    \eta(\theta) &= \frac{\theta^2}{\theta^2 + \theta_\text{maj}\theta_\text{min}},  \label{eqn:eta} \\
    S_\nu &= \frac{h\nu}{k} \frac{1}{e^\frac{h\nu}{kT} - 1} \\ 
    \tau_\nu &= \sum_t \tau^t_\nu, \\
    \tau^t_\nu &= \frac{c^2}{8\pi \nu^2} 
    N_\text{tot} \frac{ A^t_{ul} g^t_u}{Q(T_\text{ex})}  e^{-\frac{E^t_l}{k T_\text{ex}}} (1 - e^{-\frac{h\nu^t}{k T_\text{ex}}}) \phi^t_\nu,  \\
    \phi^t_\nu &= \frac{1}{\sqrt{2\pi}\sigma^t} \exp \left[ -\frac{1}{2} \left( \frac{\nu - \delta \nu^t}{\sigma^t}\right)^2  \right], \\
    \delta \nu^t &= \left( 1 - \frac{v_\text{offset}}{c} \right) \nu^t, \\
    \sigma^t &= \frac{1}{2 \sqrt{2 \ln 2}} \frac{\Delta v}{c} \delta \nu_t, 
\end{align}
}
where $J^{bg}_\nu$ is the background intensity, $\theta_\text{maj,min}$ is the major (minor) axis of the synthesis beam, $c$ is the speed of light and $k$ is the Boltzmann constant. The energy of the lower state $E^t_l$, the transition frequency $\nu^t$, the upper state degeneracy $g^t_u$, the Einstein coefficient $A^t_\text{ul}$, and the partition function $Q(T_\text{ex})$ are obtained from the spectroscopic database mentioned in Section \ref{sec:input}.
\par
The spectral line model has five fitting parameters, i.e. the source size $\theta$, the excitation temperature $T_\text{ex}$, the column density $N_\text{tot}$, the velocity width $\Delta v$, and the velocity offset $v_\text{offset}$.  According to Equation \ref{eqn:eta}, the fitted value of $\theta$ depends on the synthesis beam, which varies across different observations, while we expect that the output ranges of our neural network are unvaried. Therefore, we chose to fit the filling factor 
$\eta$ rather than the source size. The fitting parameters and their corresponding bounds are summarized in Table \ref{tab:params}.
\par
In addition, we follow the approach in \cite{2017A&A...598A...7M} to account for the instrumental resolution effect. The final output model spectrum is obtained by computing the following integral:
\begin{equation} 
    J'(\nu) = \frac{1}{\Delta \nu_\text{c}} \int^{\nu + \Delta \nu_\text{c}/2}_{\nu - \Delta \nu_\text{c}/2} J(\nu') \, d\nu', \label{eqn:slm_last}
\end{equation}
where $\Delta \nu_\text{c}$ is the channel width. The trapezoidal
rule is used to compute the above integral.
\par
We implement Equations \ref{eqn:slm_first} to \ref{eqn:slm_last} in Python and accelerate the code using Numba, a just-in-time compiler for Python.

\subsubsection{The fitting loss}
This work adopts the peak-matching loss function developed by \cite{2025ApJS..277...21Q} to measure the similarity between the model and observed spectra. This function is also used as the optimization objective for the baseline fitting tasks described in Section \ref{sec:baseline}. \cite{2025ApJS..277...21Q} demonstrated that the peak-matching loss function can lead to reasonable fitting results in the line-rich regions such as SgrB2(N), significantly more robust than the $\chi^2$ function. The peak-matching loss function can be written as
\begin{equation}
    \ell_\text{PM} = \ell_\text{MAE} + \ell_\text{peak}, \label{eqn:fitting_loss}
\end{equation}
where $\ell_\text{MAE}$ is the mean absolute error and $\ell_\text{peak}$ measures the degree to which the peaks in the model and observed spectra match. The calculation of $\ell_\text{peak}$ is complicated, and readers are referred to \cite{2025ApJS..277...21Q} for details.

\subsection{Training and testing data} \label{sec:data}
This work uses observed line cubes as training and testing data, which are summarized in Table \ref{tab:cubes}. Most data cubes are from ALMA projects, observed at bands 3, 6 and 7. Our sample includes a variety of target types, such as low- and high-mass star-forming regions, as well as ultra-compact \textsc{H ii} (UC \textsc{H ii}) region candidates.
\par
The ALMA Three-millimeter Observations of Massive Star-forming regions (ATOMS) project is a band-3 spectral line survey targeting 146 active star-forming regions across the Galactic plane \citep{2020MNRAS.496.2790L}. The sample includes hot molecular cores and UC \textsc{H ii} region candidates, with angular resolutions ranging from 1.3$''$ to 2.7$''$. The data cubes were produced by combining observations of ALMA 7-m array (ACA) and 12-m-array, using CASA version 5.6 \citep{2020MNRAS.496.2790L}. The 146 sources are divided into training and testing after preprocessing in Section \ref{sec:preprocessing}.
\par
IRAS16293–2422 is a binary system comprising two typical low-mass protostars, IRAS16293 A and IRAS16293 B, located at a distance of approximately 141 pc \citep{2018A&A...614A..20D}. The two components exhibit distinct spectral intensities and line widths. When angular resolution permits, they are treated as separate sources for the analysis. We retrieved the data cubes from five archival ALMA projects, processed using the standard pipeline and CASA versions at the time of each observation \citep{2022PASP..134k4501C}. The data cover ALMA bands 3, 6, and 7, with angular resolutions ranging from 0.5$''$, 1.0$''$, and 5.9$''$, respectively.
\par
The high-mass star-forming complex G327.3–0.6 (hereafter G327), with a luminosity of 2$\times$10$^5$ L$_\odot$ and at a distance of 3.3 kpc, is best known as one of the brightest and chemically richest hot molecular cores in our vicinity\citep{2006A&A...454L..91W}. For this source, we used data cubes from three ALMA projects, all observed with the 12-m array. The angular resolutions vary across frequency bands: 0.1$''$ and 1.9$''$ in ALMA band 6 and 1.1$''$ in band 7. Calibration and imaging, including self-calibration where applicable, were performed using the CASA version 5.6.1.
\par
We also include a line survey of Orion-KL obtained from combined Submillimeter Array (SMA) interferometric and IRAM 30 m single-dish observations \citep{2015A&A...581A..71F}. It covers 4 GHz at band 6, with a resolution of 4.9$''$. \citet{2015A&A...581A..71F} converted the 30 m single-dish data into visibilities, and combined them with the SMA data using the MIRIAD package UVMODEL task \footnote{\url{https://www.cfa.harvard.edu/sma/miriad/manuals/SMAuguide/smauserhtml/uvmodel.html}}.

\begin{table*}
\centering
\caption{Summary of the observational data used in this work.}
\label{tab:cubes}
\begin{tabular}{lccccccc}
    \hline
    \hline
    Dataset & ALMA project ID & Beam & Freq. range & Freq. coverage$^a$ & $\delta f$$^b$ & $\langle \sigma^\text{global}_\text{RMS} \rangle$$^c$& $N_\text{fit}$$^d$\\
    && ($''$) & (GHz) & (GHz) & (kHz) & (K) & \\
    \hline
    \multicolumn{8}{c}{Training} \\
    \hline
    ATOMS-80-band3-2as & 2019.1.00685.S & 1.3 - 2.7 & 97.57 - 101.39 & 3.74 & 488 & 0.129 & 67098 \\
    IRAS16293 A-band6-0.5as & 2012.1.00712.S & 0.5 & 231.01 - 250.72 &  1.81 & 122 & 0.250 & 5721 \\
    IRAS16293 B-band6-0.5as & 2012.1.00712.S & 0.5 & 231.01 - 250.72 &  1.81 & 122 & 0.250 & 9350 \\
    IRAS16293 A-band7-0.4as & 2018.1.01496.S & 0.4 & 281.68 - 283.88 &  0.94 & 244 & 1.652 & 5144 \\
    IRAS16293 B-band7-0.4as & 2018.1.01496.S & 0.4 & 281.68 - 283.88 &  0.94 & 244 & 1.652 & 4820 \\
    IRAS16293 A-band6-0.5as & 2021.1.01164.S & 0.5 & 232.05 - 240.80 &  8.74 & 122 & 1.024 & 9350 \\
    IRAS16293 B-band6-0.5as & 2021.1.01164.S & 0.5 & 232.05 - 240.80 &  8.74 & 122 & 1.024 & 13403 \\
    IRAS16293-band6-5.9as & 2017.1.00108.S & 5.9 & 232.06 - 233.93 &  1.87 & 488 & 0.011 & 5611 \\
    G327-band6-0.1as & 2022.1.01354.S & 0.1 & 216.70 - 236.07 &  7.50 & 977 & 1.864 & 11766\\
    G327-band6-1.9as & 2016.1.00168.S & 1.9 & 215.56 - 231.32 &  5.12 & 488,977 & 0.048 & 5702\\
    G327-band7-1.1as & 2024.1.00653.S & 1.1 & 279.44 - 294.37 &  5.84 & 122,977 & 0.327 & 9462\\
    \hline
    \multicolumn{8}{c}{Testing} \\
    \hline
    ATOMS-20-band3-2as & 2019.1.00685.S & 1.3 - 2.7 & 97.57 - 101.39 &  3.74 & 488 & 0.129 & 16240 \\ 
    IRAS16293 A-band3-1.0as & 2015.1.01193.S & 1.0 &  86.58 - 101.57 & 1.64 & 244 & 0.399 & 3704 \\
    IRAS16293 B-band3-1.0as & 2015.1.01193.S & 1.0 &  86.58 - 101.57 & 1.64 & 244 & 0.399 & 4755 \\
    Orion KL-band6-4.9as & - & 4.9 & 221.15 - 233.42 & 4.46 & 880,921 & 0.080 & 8044 \\
    \hline
\end{tabular}
\tablecomments{$^a$The frequency coverage is derived by summing the bandwidth of all spectral windows. \\ $^b$This is the resolution of frequency channels. \\ $^c$The estimation of the global RMS noise is introduced in Section \ref{sec:preprocessing}. The average is taken over all spectral windows. \\ $^d$This is the number of fitting results for evaluation as described in Section \ref{sec:baseline}.}
\end{table*}

\begin{table*}
    \label{tab:mols}
    \centering
    \caption{Molecular species used for neural network training and testing.}
    \begin{tabular}{p{.13\textwidth}p{.13\textwidth}p{.13\textwidth}p{.13\textwidth}p{.13\textwidth}p{.13\textwidth}p{.13\textwidth}p{.13\textwidth}}
    \hline \hline
    \multicolumn{7}{c}{Mol-1980} \\
    \hline
    C$_{2}$H & C$_{2}$H$_{2}$ & C$_{2}$H$_{3}$CN & C$_{2}$H$_{5}$CN & C$_{2}$H$_{5}$OH & C$_{3}$N & C$_{4}$H \\
    CH & CH+ & CH$_{2}$NH & CH$_{3}$CCH & CH$_{3}$CHO & CH$_{3}$CN & CH$_{3}$NH$_{2}$ \\
    CH$_{3}$OCH$_{3}$ & CH$_{3}$OCHO & CH$_{3}$OH & CH$_{3}$SH & CN & CO & CS \\
    H$_{2}$CCO & H$_{2}$CO & H$_{2}$CS & H$_{2}$O & H$_{2}$S & HC(O)NH$_{2}$ & HC$_{3}$N \\
    HC$_{5}$N & HC$_{7}$N & HC$_{9}$N & HCN & HCO & HCO+ & HCOOH \\
    HNC & HNCO & HNCS & HNO & N$_{2}$H+ & NH$_{2}$CN & NH$_{3}$ \\
    NO & NS & OCS & OH & SO & SO$_{2}$ & SiO \\
    SiS &  &  &  &  &  &  \\
    \hline
    \multicolumn{7}{c}{Mol-2010} \\
    \hline
    AlCl & AlF & AlNC & AlO & AlOH & C$_{2}$H & C$_{2}$H$_{2}$ \\
    C$_{2}$H$_{3}$CHO & C$_{2}$H$_{4}$ & C$_{2}$H$_{5}$CHO & C$_{2}$H$_{5}$OCHO & C$_{2}$H$_{5}$OH & C$_{2}$O & C$_{2}$S \\
    C$_{3}$ & C$_{3}$H & C$_{3}$H$_{7}$CN & C$_{3}$N & C$_{3}$N- & C$_{3}$O & C$_{3}$S \\
    C$_{4}$H & C$_{4}$H- & C$_{5}$H & C$_{5}$N & C$_{5}$N- & C$_{5}$S & C$_{6}$H \\
    C$_{6}$H- & C$_{7}$H & C$_{8}$H & C$_{8}$H- & CCP & CF+ & CH$_{2}$ \\
    CH$_{2}$CHOH & CH$_{2}$NH & CH$_{3}$C$_{3}$N & CH$_{3}$C$_{4}$H & CH$_{3}$C$_{5}$N & CH$_{3}$C$_{6}$H & CH$_{3}$CCH \\
    CH$_{3}$CHO & CH$_{3}$CN & CH$_{3}$COCH$_{3}$ & CH$_{3}$COOH & CH$_{3}$NC & CH$_{3}$NH$_{2}$ & CH$_{3}$OCH$_{3}$ \\
    CH$_{3}$OCHO & CH$_{3}$OH & CH$_{3}$SH & CH$_{4}$ & CN- & CO+ & CP \\
    H$_{2}$ & H$_{2}$CCCHCN & H$_{2}$CCN & H$_{2}$CCO & H$_{2}$CN & H$_{2}$CNH & H$_{2}$CO \\
    H$_{2}$COH+ & H$_{2}$CS & H$_{2}$Cl+ & H$_{2}$NCH$_{2}$CN & H$_{2}$O & H$_{2}$O+ & H$_{2}$S \\
    H$_{3}$+ & H$_{3}$O+ & HC(O)CN & HC(O)NH$_{2}$ & HC$_{2}$CHO & HC$_{2}$N & HC$_{3}$N \\
    HC$_{3}$NH+ & HC$_{5}$N & HC$_{7}$N & HC$_{9}$N & HCCNC & HCNH+ & HCNO \\
    HCOCH$_{2}$OH & HCP & HCS+ & HCl & HF & HNCCC & HOC+ \\
    HOCN & HOCO+ & HSCN & KCN & KCl & MgCN & MgNC \\
    N$_{2}$H+ & N$_{2}$O & NH & NH$_{2}$ & NH$_{2}$CN & NH$_{3}$ & NaCN \\
    NaCl & O$_{2}$ & OH+ & PH$_{3}$ & PN & PO & SO+ \\
    SO$_{2}$ & SiC & SiC$_{2}$ & SiC$_{4}$ & SiCN & SiN & SiNC \\
    aGg'-(CH$_{2}$OH)$_{2}$ & c-C$_{2}$H$_{4}$O & c-C$_{3}$H$_{2}$ & c-SiCCC & gGg'-(CH$_{2}$OH)$_{2}$ & l-C$_{3}$H$_{2}$ & l-C$_{4}$H$_{2}$ \\
    l-C$_{6}$H$_{2}$ & l-H$_{2}$CCCO & l-HCCCCN & l-SiCCC &  &  &  \\
    \hline
    \end{tabular}
\end{table*}

\subsubsection{Data preprocessing} \label{sec:preprocessing}
Our data preprocessing has two primary purposes:
\begin{enumerate}
    \item We discard pixels that contain too few peaks. Performing spectral line fitting on such pixels cannot obtain meaningful results, and therefore hindering neural network training.
    \item We estimate the RMS noise of each spectrum, a quantity required by both the line-fitting routine and the neural network.
\end{enumerate}
The procedure consists of the following steps:
\begin{enumerate}
    \item For each line cube, we randomly select at most 10,000 pixels and apply a sigma clipping algorithm to all frequency channels of the selected pixels to estimate the global RMS noise $\sigma^\text{global}_\text{RMS}$. We use \textit{sigma\_clip} implemented in \textit{astropy}. 
    \item We employ \textit{find\_peaks} implemented in \textit{scipy} to count the number of peaks for each spectral window, with a prominence threshold of $4\sigma^\text{global}_\text{RMS}$. We then estimate the mean number of peaks per spectral window for each pixel and discard pixels whose mean is below $N_\text{cut}$.
    \item For each pixel, we use the spectrum to estimate the local RMS noise $\sigma^\text{local}_\text{RMS}$ for each spectral window. 
    \item We count the number of peaks, with a prominence threshold of $6\sigma^\text{local}_\text{RMS}$, for each spectral window, and discard the pixels if the mean number of such peaks per spectral window is below $N_\text{cut}$. This step removes regions that have a very high RMS noise but few peaks.
\end{enumerate}
Starting with $N_\text{cut} = 3$, we visually check the processed cubes. If the cubes still contain a large number of noisy pixels, we will repeat the preprocessing pipeline with larger $N_\text{cut}$.
\par
We correct the local standard of rest (LSR) velocity for cubes to ensure that the velocity offsets of the species are within the fitting range given in Table \ref{tab:params}. For the ATOMS data, we adopt the values estimated by \citet{2020MNRAS.496.2790L}. For IRAS16293, \citet{2024A&A...686A..59N} assumed $v_\text{LSR} = 2.7$ km/s. We neglect such small LSR velocity and apply no velocity correction to all the IRAS16293 data. In terms of G327, the LSR velocity is roughly between -43.2 to -46.0 km/s \citep{2017A&A...602A..70L,2020ApJ...904..168B} and we adopt $v_\text{LSR} = -45.0$ km/s in this work. \citet{2015A&A...581A..71F} corrected the Orion-KL data assuming $v_\text{LSR} = 7$ km/s, and no further correction is applied in this work.
\par
In addition, continuum data are taken into count, which serve as an input feature for our neural network (see Section \ref{sec:input}) and are used to compute the model spectrum (see Equation \ref{eqn:slm_first}).
\par
For the ATOMS data, we exclude data cubes with fewer than 20 effective pixels. The remaining sources are divided into training and testing sets in an 8:2 ratio, denoted as ATOMS-80-band3-2as and ATOMS-20-band3-2as.

\subsubsection{Species} \label{sec:species}
This work uses observational data to train the neural network and performs matching between the observed and model spectra during training. Not all species in the database can be fitted and identified from the observed spectra. If there are no matches between the model and observed spectral lines, the fitting parameters may become arbitrary. Such cases can negatively affect network training. Therefore, it is essential to select the appropriate species for training.
\par
In this work, we make an assumption that molecules detected earlier are more likely to yield meaningful fitting results. The fact that these molecules were detected using early instruments with limited sensitivity and resolution suggests that they are widely present in the interstellar medium.
\par
Based on the above assumption, this work defines two sets of molecular species, named Mol-1980 and Mol-2010. The former includes species first detected before or in 1980, while the latter includes those detected between 1980 and 2010. We use \textsc{astromol} \citep{2021zndo...5046939M,2022ApJS..259...30M} to determine the year each molecule was first detected, and list the molecules in Table \ref{tab:mols}. Isotopologues are included in the corresponding species set regardless of their detection date or whether they have been detected at all. Hyperfine structure lines are excluded. We use only the species in Mol-1980 for training, and those in Mol-2010 to evaluate the generalization ability of our neural network.
\par
For each observed spectrum, we apply an additional algorithm to select species for fitting. Given an observed spectrum and a set of transition frequencies, we search for a peak in the spectrum within the velocity offset range given in Table \ref{tab:params}, and only fit species that have at least one corresponding peak. 

\begin{figure}
	\includegraphics[width=.9\columnwidth]{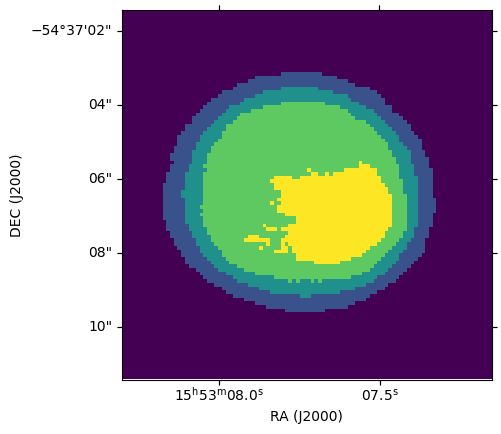}
	\centering
    \caption{Pixel binning for G327-band6-1.9as. Different colors correspond to pixels in different bins. This binning is used for data sampling, as described in Section \ref{sec:sampling}.}
    \label{fig:binning}
\end{figure}

\subsubsection{Data sampling} \label{sec:sampling}
This section explains how the data are sampled during training and producing baseline data in Section \ref{sec:baseline}. To generate a data sample, we first randomly select a dataset. The datasets of the ATOMS data contain multiple sources, while other datasets only contain a single source. If an ATOMS dataset is selected, we randomly select a source from it. Next, we randomly choose a spectrum from a pixel in the selected source using the approach described by the subsequent paragraph. Finally, we randomly select one species from the spectroscopic database using the method described in Section \ref{sec:species}.
\par
For most observations, line-rich pixels are concentrated within a compact region. Consequently, if we randomly select a pixel from a line cube, the pixel may not come from a line-rich region and may contain only a few lines. To resolve the issue, we bin the pixels in the line cubes of all spectral windows based on the total number of peaks estimated assuming a prominence threshold of $4\sigma^\text{global}_\text{RMS}$. The method to find the peaks is described in Section \ref{sec:architecture}. For all observations, we use five evenly spaced bins between the minimum and maximum number of peaks. When a spectrum is needed, we first randomly select a bin and then randomly select a pixel within the bin. As an example, Figure \ref{fig:binning} illustrates the pixel binning for G327-band7-1.9as.

\subsubsection{Data augmentation} \label{sec:augment}
Data augmentation is a technique that reduces overfitting and improves generalization by applying random transformations to the training data \citep[e.g.][]{wang2017effectiveness,shorten2019survey}. In this study, data augmentation is applied during training using the following strategies:
\begin{enumerate}
    \item Shift the observed spectrum by a velocity offset uniformly sampled between -5 to 5 km/s. The species fitted using the same observed spectrum generally have similar velocity offsets. Randomly shifting the observed spectrum improves the diversity of the fitted velocity offsets.
    \item Multiply $\sigma^\text{global}_\text{RMS}$ of the observed spectrum by a factor sampled uniformly from the range 0.85 to 1.25. This quantity is shared across all pixels in the same data cube. Random sampling improves its diversity. 
    \item Smooth the observed spectrum with 20\% probability using a top-hat kernel. The kernel size $K$ is randomly selected from 2, 4 and 8 frequency channels. The RMS noise of the spectrum is then divided by $\sqrt{K}$. This operation also increases the variability of the RMS noise.
    \item Multiply the continuum temperature by a factor sampled randomly from the range 0.5 to 2. This accounts for uncertainties in the estimation of continuum temperature.
    \item Multiply the beam size by a factor sampled randomly from the range 0.5 to 2. The beam size remains the same across all pixels in the same data cube. This sampling increases the variability of the beam size values.
\end{enumerate}

\subsection{Network training} \label{sec:training}
We frame the training of our neural network as a single-step reinforcement learning problem. At each training step, the neural network takes the molecular spectroscopic data and the observed spectrum as input, and samples some fitting parameters. A reward value is then assigned by comparing the resulting model spectrum with the observed spectrum. Samples that result in a better match to the observed spectrum receive higher rewards. These reward values serve as training signals for updating the network, a method commonly known as policy gradient \citep[e.g.][]{williams1992simple}.
\par
Denote the neural network as $P_\theta(\textbf{a}|\textbf{d}_\text{trans}, \textbf{d}_\text{spec})$, representing the conditional probability of the fitting parameters $\textbf{a} = (\eta, T_\text{ex}, N_\text{tot}, \Delta v, v_\text{offset})$ given the molecular spectroscopic data $\textbf{d}_\text{trans}$ and the observed spectrum $\textbf{d}_\text{spec}$ with trainable weights $\phi$. The training loss is computed as
\begin{align}
    \mathcal{L} &= \mathcal{L}_\text{PG} + \mathcal{L}_\text{RB} + \lambda_\text{ER} \mathcal{L}_\text{ER},  \\
    \mathcal{L}_\text{PG} &= -\mathbb{E}_{\textbf{a} \sim P_\phi} \left[ R(\textbf{a}) \log P_\phi(\textbf{a}|\textbf{d}_\text{trans}, \textbf{d}_\text{spec}) \right], \label{eqn:l_pg} \\
    \mathcal{L}_\text{ER} &= \mathbb{E}_{\textbf{a} \sim P_\phi} \left[ \log P_\phi(\textbf{a}|\textbf{d}_\text{trans}, \textbf{d}_\text{spec}) \right],  \label{eqn:l_er} \\
    \mathcal{L}_\text{RB} &= -\mathbb{E}_{\textbf{a} \sim \text{RB}} \left[ \log P_\phi(\textbf{a}|\textbf{d}_\text{trans}, \textbf{d}_\text{spec}) \right].  \label{eqn:l_rb}
\end{align}
\par
The first term $\mathcal{L}_\text{PG}$ is the standard policy gradient, with the reward term defined as:
\begin{equation}
    R(\textbf{a})=
    \begin{cases}
    1, \, l_\text{PM}(\textbf{a}) < l^\text{med}_\text{PM}, \\
    0, \, l_\text{PM}(\textbf{a}) \geq l^\text{med}_\text{PM} \text{ and } l_\text{peak}(\textbf{a}) \neq 0,  \\
    -0.5, \, l_\text{PM}(\textbf{a}) \geq l^\text{med}_\text{PM} \text{ and } l_\text{peak}(\textbf{a}) = 0,
    \end{cases}
\end{equation}
where $l_\text{PM}$ and $l_\text{peak}$ are defined in Equation \ref{eqn:fitting_loss}. We compute the rewards based on a relative criterion. For each $\textbf{d}_\text{trans}$ and $\textbf{d}_\text{spec}$, we draw $N_\text{draw} = 64$ samples and compute the median score $ l^\text{med}_\text{PM}$ to determine the rewards. In addition, when $l_\text{peak} =0$, it implies that no peaks are detected in the model spectrum. These cases are unexpected and are penalized with a negative reward.
\par
The second term $\mathcal{L}_\text{ER}$ is known as the entropy regularization. This term stabilizes the training, preventing the neural network from becoming overly confident in its predictions too early. The factor $\lambda_\text{ER}$ controls the strength of the entropy regularization, and we suggest $\lambda_\text{ER}=0.05$ after several experiments.
\par
Finally, the third term $\mathcal{L}_\text{RB}$ corresponds to the prioritized replay buffer, a technique we introduce to enhance training. A similar idea was presented by \cite{horgan2018distributed}. The prioritized replay buffer stores the best-fitting samples from previous iterations, which is implemented using a priority queue, a standard data structure in computer science. When adding an item to a full priority queue, the lowest-priority item is automatically removed. In this work, samples with lower $l_\text{PM}$ values are assigned higher priority. Therefore, after many iterations, samples with lower $l_\text{PM}$ values remain in the priority queue.
\par
We construct the prioritized replay buffer by randomly selecting observed spectra and species for fitting. Each entry in the replay buffer is identified by four attributes: the dataset name, pixel index, target molecule, and augmentation parameters (see Section \ref{sec:augment}). These entries maintain a priority queue that stores $l_\text{PM}$ values and their associated fitting parameters. All priority queues have a fixed size of 32. For each training dataset except ATOMS-80-band3-2as, we add 16 randomly chosen pixels from each line number bin to the buffer. For ATOMS-80-band3-2as, we instead add 5 pixels per line number bin from each source. These entries are prepared before training. During training, in addition to computing Equation \ref{eqn:l_er}, we introduce an extra step to update the prioritized replay buffer, which is found to improve training performance in practice.
\par
Every training step proceeds as follows:
\begin{enumerate}
    \item Sample $N_\text{batch}/2$ observed spectra from the training datasets, each associated with a target molecule (or isotopologue) for fitting.  Feed the input data through the neural network to generate $N_\text{draw}$ sets of fitting parameters and interact with the spectral line fitting module to compute Equations \ref{eqn:l_pg} and \ref{eqn:l_er}.
    \item Sample $N_\text{batch}/2$ items from the prioritized replay buffer. Feed the input data through the neural network to generate $N_\text{draw}$ sets of fitting parameters for each item. Update the priority queues using these new samples and compute Equation \ref{eqn:l_er} using the remaining samples in the queues.
    \item Perform backward propagation and update the neural network weights.
    \item Sample $4N_\text{batch}$ items from the prioritized replay buffer. For each item, generate $N_\text{draw}$ parameter samples using the neural network and update the corresponding priority queues.
\end{enumerate}
Here, $N_\text{batch} = 256$ is the total batch size. 
\par
Our neural network is trained using 32 V100 GPUs with 256 CPU cores. We adopt a AdamW optimizer \citep{loshchilov2017decoupled} with a weight decay of 0.04 and a learning rate scheduler proposed by \citep{smith2019super}. The learning rate first increases from $7.5 \times 10^{-7}$ to $7.5 \times 10^{-5}$ for 4000 steps and decreases to $7.5 \times 10^{-9}$ for 31000 steps according to a cosine function. The training takes 3 - 4 days.

\begin{table*}
\centering
\caption{Summary of the optimization algorithms used in inference.}
\label{tab:optimizers}
\begin{tabular}{cccc}
    \hline
    \hline
    Method & Scope & Gradient & Reference \\
    \hline
    Nelder-Mead  & Local  & No & \citet{gao2012implementing} \\
    COBYLA       & Local  & No & \citet{powell1994direct,Zhang_2023} \\
    COBYQA       & Local  & No & \citet{rago_thesis} \\
    L-BFGS-B     & Local  & Yes & \citet{byrd1995limited} \\
    TNC          & Local  & Yes & \citet{nash1984newton} \\
    SLSQP        & Local  & Yes & \citet{kraft1988software} \\
    PSO          & Global & No & \citet{wang2018particle} \\
    \hline
\end{tabular}
\tablecomments{All local optimizers are implemented in \textsc{scipy} \citep{2020SciPy-NMeth}.}
\end{table*}

\begin{figure*}
	\includegraphics[width=\textwidth]{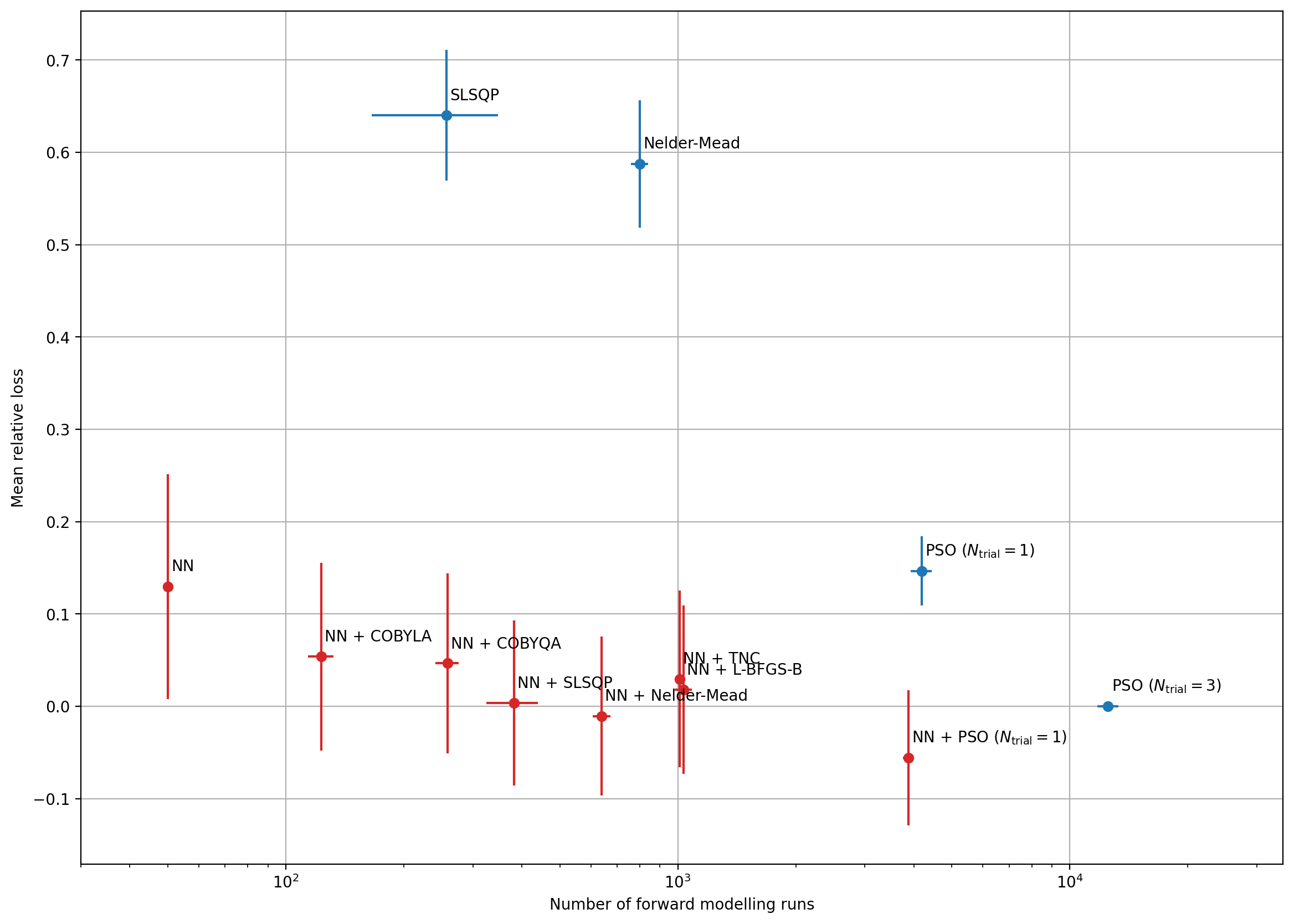}
	\centering
    \caption{Relation between the mean relative loss and the number of forward modeling runs using different fitting methods. Each dot shows the average metric over all datasets, and the error bars represent the standard deviation over the datasets. The mean relative loss is defined in Section \ref{sec:metric}. The datasets are described in Section \ref{sec:data} and summarized in Table \ref{tab:cubes}. Blue dots correspond to results based on traditional optimization methods. Nelder-Mead and SLSQP are classical local optimizers, whereas particle swarm optimization (PSO) is a global optimization algorithm. Red dots show results where a neural network generates the initial guess for the local optimizers. For the methods based on local optimization, the initial guess is given by the best parameters from $N_\text{init} = 50$ sampled points. The methods to produce these results are introduced in Sections \ref{sec:inference} and \ref{sec:baseline}.}
    \label{fig:err_nfev_1}
\end{figure*}

\begin{figure*}
	\includegraphics[width=\textwidth]{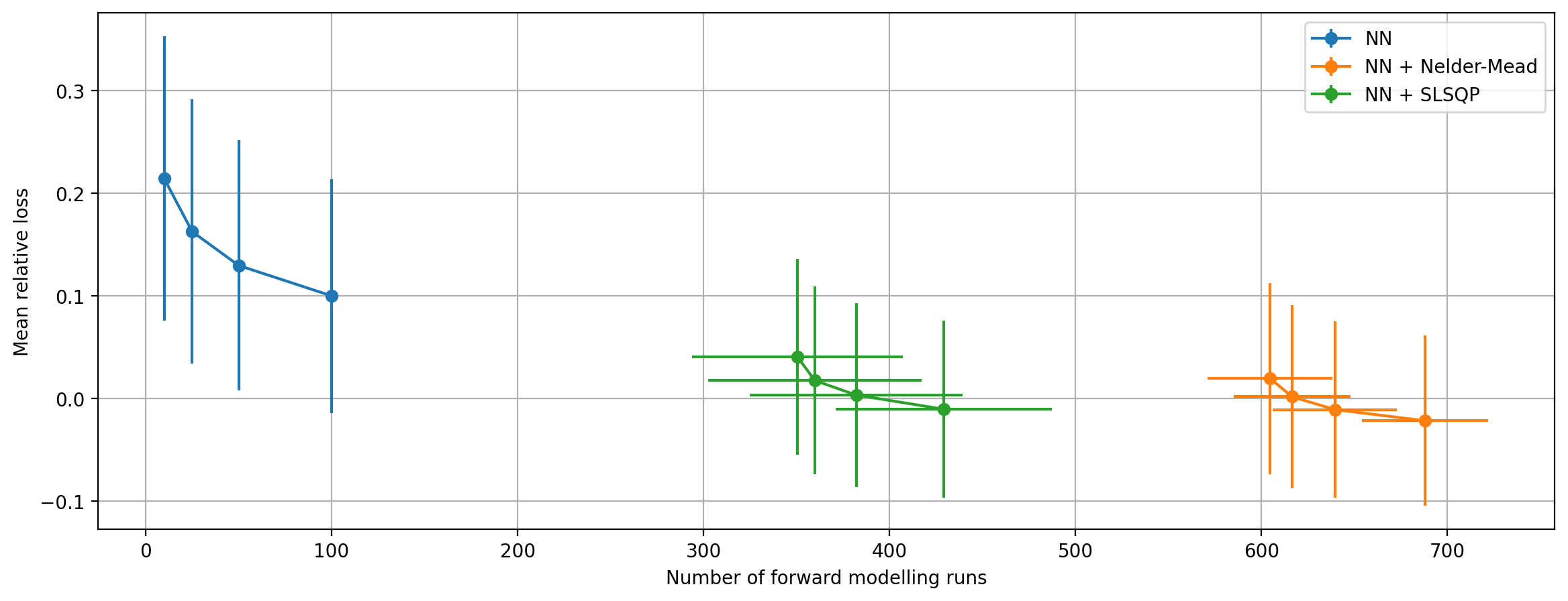}
	\centering
    \caption{Relation between the mean relative loss and the number of forward modeling runs with different number of initial points generated by the neural network. For each method, from left to right, the dots show the average metric over all datasets with $N_\text{init}=10, 25, 50, 100$. The error bars represent the standard deviation over the datasets. The mean relative loss is defined in Section \ref{sec:metric}. The datasets are described in Section \ref{sec:data} and summarized in Table \ref{tab:cubes}.}
    \label{fig:err_nfev_2}
\end{figure*}

\begin{figure*}
	\includegraphics[width=\textwidth]{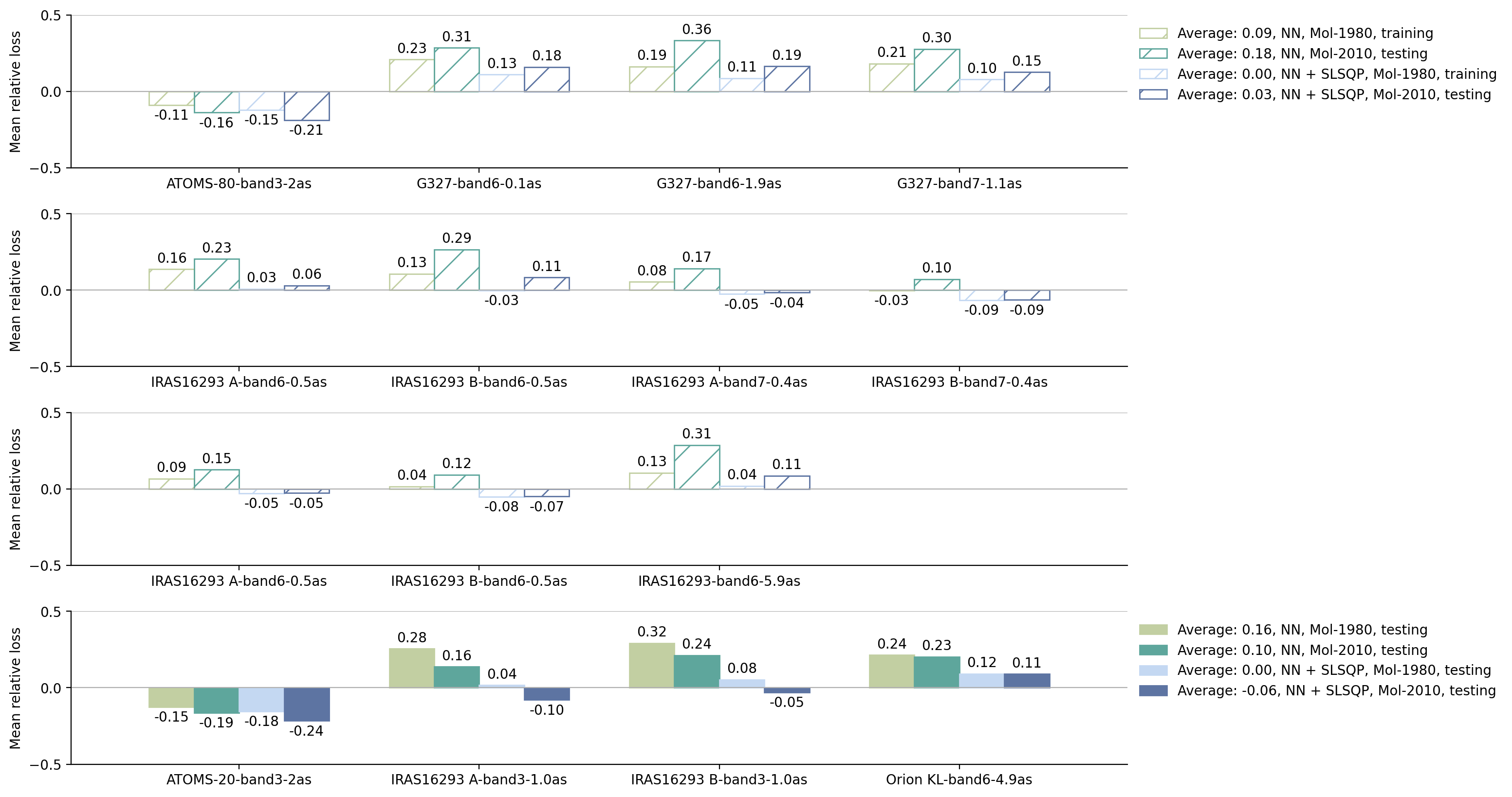}
	\centering
    \caption{Mean relative loss of neural network-based methods across individual datasets. For each dataset, we present the results from using the neural network alone and from combining the neural network with the SLSQP optimization algorithm. The results are divided into two sets based on the species for fitting, i.e. Mol-1980 and Mol-2010 as described in Section \ref{sec:species}. While Mol-1980 is used for training, Mol-2010 is reserved for testing. The mean relative loss metric is defined in Section \ref{sec:metric}, and the inference method is introduced in Section \ref{sec:inference}.}
    \label{fig:err_ds}
\end{figure*}

\subsection{Inference methods} \label{sec:inference}
After training, the neural network can suggest favorable fitting parameters, which are then passed to a local optimization algorithm to further improve the fitting results. The proposed inference method consists of the following steps:
\begin{enumerate}
    \item Sample $N_\text{init}$ sets of parameters using the neural network and find the best parameters with lowest fitting loss.
    \item Use the best parameters as an initial guess for a subsequent fitting process performed by an optimization algorithm.
\end{enumerate}
This work assesses the performance of six local optimization algorithms from \textit{scipy} \citep{2020SciPy-NMeth}, summarized in Table \ref{tab:optimizers}. To prevent parameter divergence, we restrict our selection to algorithms that support bounded parameter ranges. For gradient based optimizers, i.e. L-BFGS-B, TNC, and SLSQP, a three point numerical differentiation formula is used to compute the gradient. We test different $N_\text{init}$ and present the results in Section \ref{sec:results}. We also evaluate particle swarm optimization (PSO) \citep[e.g.][]{wang2018particle}, which can be combined with the neural network by generating all initial particle positions using the outputs of the neural network.
\par
The role of our neural network is to provide an initial guess, and the subsequent fitting process can in principle be applied with any loss function. While the peak matching loss function is employed during training and evaluation, the proposed inference method can also be applied to $\chi^2$ fitting. In Section \ref{sec:app}, we demonstrate that both the peak matching and $\chi^2$ functions can lead to reasonable but different fitting results.

\subsection{Producing evaluation data} \label{sec:baseline}
For evaluation, we compare the proposed neural network-based method with traditional optimization algorithms. First, we construct a sample of observed spectra for fitting using the sampling method described in Section \ref{sec:sampling}. For each dataset, we randomly select five spectra from each line number bin. The ATOMS data, i.e. ATOMS-80-band3-2as and ATOMS-20-band3-2as, contain multiple sources. Therefore, we use all sources in the datasets, and select only one spectrum per line number bin for each source. We then perform spectral line fitting on the selected observed spectra using different optimization algorithms. The species used for fitting are from both Mol-1980 and Mol-2010, as described in Section \ref{sec:species}. The number of fitting results for each dataset is presented in Table \ref{tab:cubes}.
\par
In this study, three optimization methods are compared with the neural network-based method. First, particle swarm PSO \citep[e.g.][]{wang2018particle} is executed to search the global minimum. The algorithm is run three times to achieve better results. Following \cite{2025ApJS..277...21Q}, we terminate the algorithm if there is no improvement over 15 consecutive iterations, with a minimum of 100 iterations. The population size is set to 28. Secondly, we use randomly initialized local optimizers to demonstrate the effectiveness of the neural network. The Nelder-Mead \citep{gao2012implementing} and SLSQP \citep{kraft1988software} algorithms are chosen as representatives of gradient-free and gradient-based methods, respectively. The best result among $N_\text{init}$ random samples within the bounds specified in Table \ref{tab:params} is selected as the initial guess for the local optimizers.

\subsection{Metrics} \label{sec:metric}
To compare the fitting results, we use a mean relative loss defined as
\begin{equation}
    \epsilon = \frac{1}{n} \sum_{i=1}^n\frac{\ell^\text{pred}_\text{PM} - \ell^\text{test}_\text{PM}}{|\ell^\text{pred}_\text{PM}| + |\ell^\text{test}_\text{PM}|},
\end{equation}
where $\ell^\text{pred}_\text{PM}$ and $\ell^\text{test}_\text{PM}$ represent the peak matching fitting loss defined in Equation \ref{eqn:fitting_loss}. The proposed metric is inspired by the symmetric mean absolute percentage error (SMAPE) \citep{nguyen2019efficient}. Since the values of the peak matching function vary substantially across different observed spectra and fitting species, we use a relative metric to ensure comparability of the fitting results. Secondly, the ground truths in the our fitting problems are unknown, and we aim to identify methods that produce lower fitting loss values. Therefore, we avoid absolute-value-based metrics. The proposed metric ranges from -1 to 1, with smaller values indicating better performance. This work adopts the results based on PSO with $N_\text{trail} = 3$ for $\ell^\text{test}_\text{PM}$, which in practice produces acceptable fitting loss values.

\begin{figure*}
	\includegraphics[width=\textwidth]{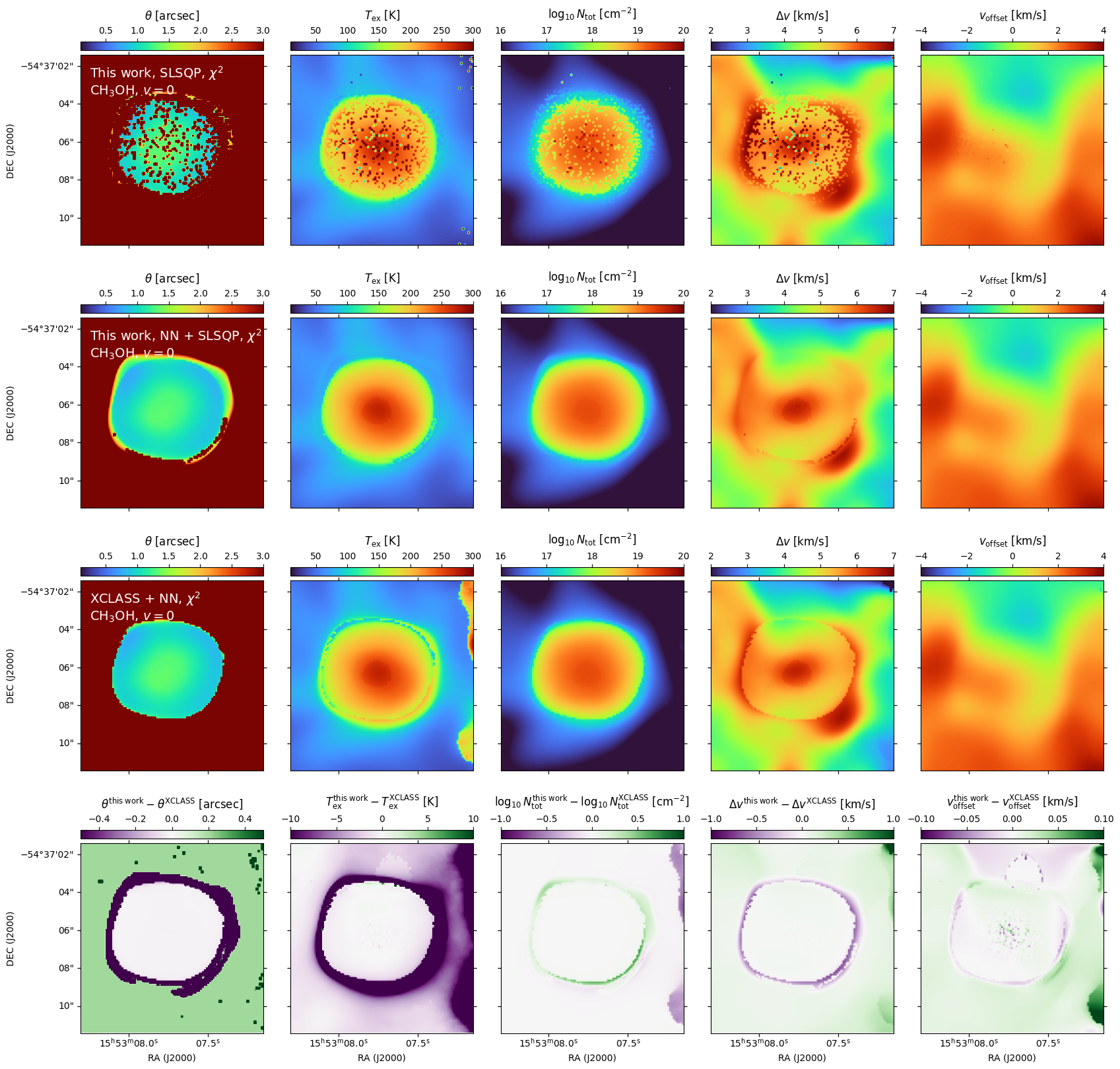}
	\centering
    \caption{Comparison of our pixel-level fitting results with \textsc{xclass}. The fitting region is from an ALMA observation of G327, covering a total bandwidth of $\sim 5$ GHz. The data are described in Section \ref{sec:data}. The fitted molecule is methanol. From left to right, the columns show the fitting results for source size, excitation temperature, column density, velocity width and velocity offset. The spectral line model and these fitting parameters are described in Section \ref{sec:sl_model}. The first and second rows display our results. The first row uses random initial guesses while the second row employ the neural network to generate initial guesses. The third row illustrates the \textsc{xclass} results. The predictions from our neural network are also used as initial guesses to generate these results. Finally, the fourth row shows the difference between our results and those based on \textsc{xclass}.}
    \label{fig:map_xclass}
\end{figure*}

\begin{figure*}
	\includegraphics[width=\textwidth]{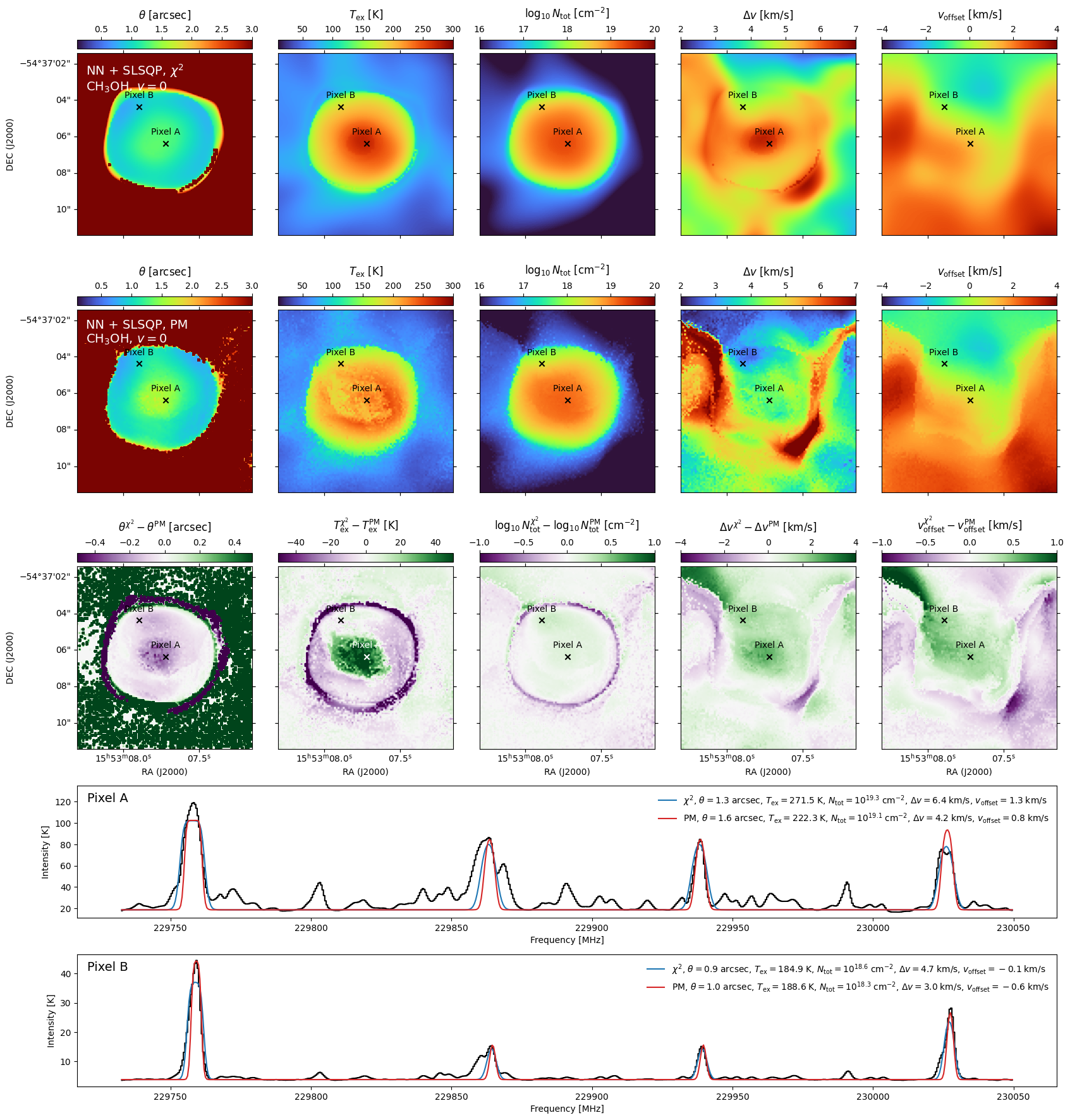}
	\centering
    \caption{Comparison of pixel-level fitting results using the $\chi^2$ and peak matching loss functions. The fitting region is from an ALMA observation of G327 at band 6, covering a total bandwidth of $\sim 5$ GHz. The data are described in Section \ref{sec:data}. The fitted molecule is methanol. From left to right, the columns show the fitting results for source size, excitation temperature, column density, velocity width and velocity offset. The spectral line model and these fitting parameters are described in Section \ref{sec:sl_model}. The first and second rows display the results using the $\chi^2$ and peak matching functions respectively, and the third row shows their difference. The fourth and fifth rows illustrate the fitted spectra for two pixels, with black lines representing the observed spectra. Only a representative frequency range is shown for clarity.}
    \label{fig:map_chi2_pm_CH3OH}
\end{figure*}

\begin{figure*}
	\includegraphics[width=\textwidth]{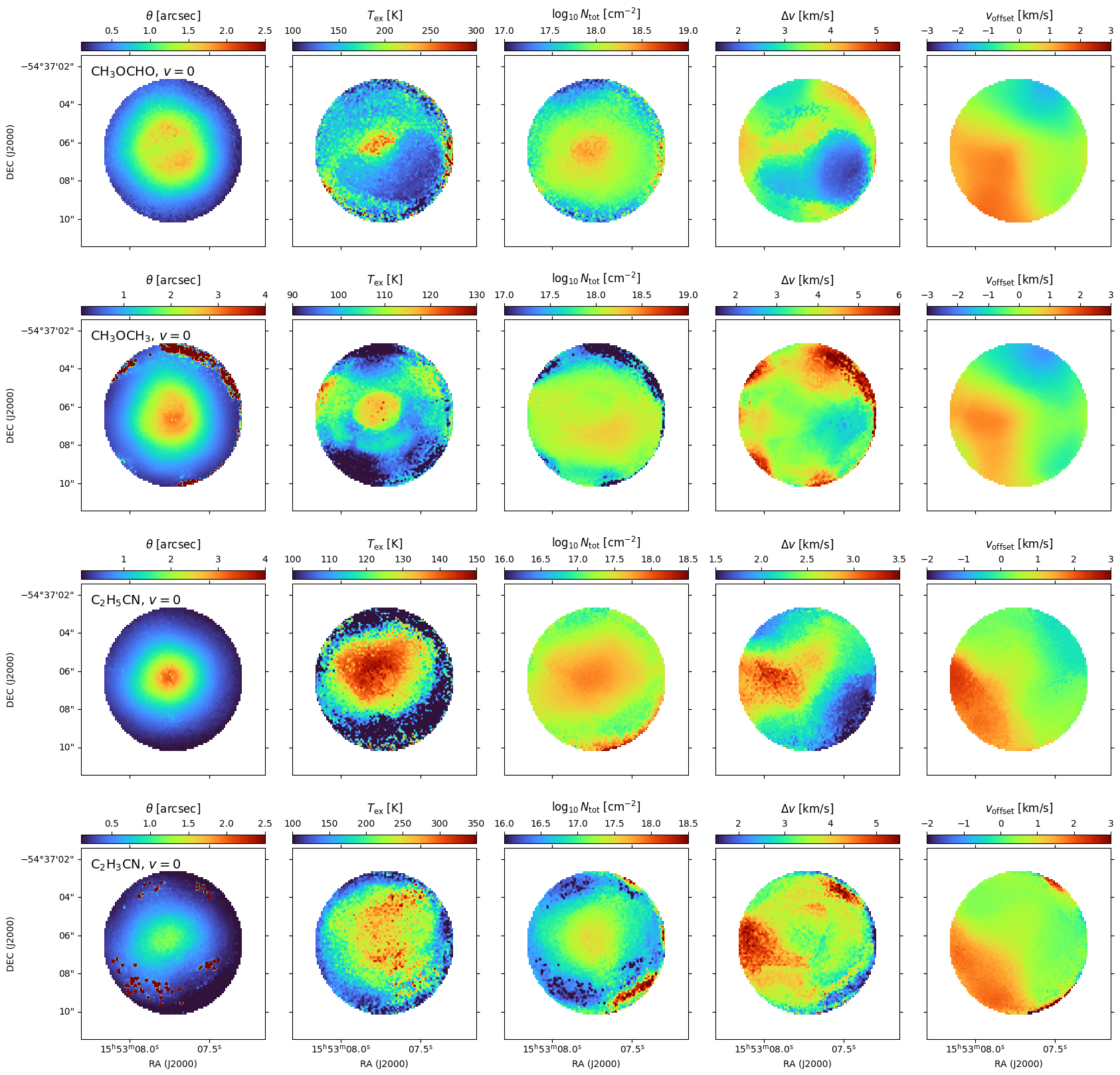}
	\centering
    \caption{Pixel-level fitting results of CH$_3$OCHO, $v=0$, CH$_3$OCH$_3$, $v=0$, C$_2$H$_5$CN, $v=0$, and C$_2$H$_3$CN, $v=0$. The fitting region is from an ALMA observation of G327 at band 6, covering a total bandwidth of $\sim 5$ GHz. The data are described in Section \ref{sec:data}. These fitting results are obtained using the SLSQP algorithm, with initial guesses generated using the neural network. The peak matching loss function is adopted in the fittings. From left to right, the columns show the fitting results for source size, excitation temperature, column density, velocity width and velocity offset. The spectral line model and these fitting parameters are described in Section \ref{sec:sl_model}.
    For clarity, we only demonstrate results in the central region, and the color scales of each row are different.}
    \label{fig:map_mols}
\end{figure*}

\section{Results} \label{sec:results}
Figure \ref{fig:err_nfev_1} illustrates the relation between the relative fitting error and the number of forward modeling runs for different fitting methods. All the results are produced with $N_\text{init}=50$. We show the average metrics over all datasets, and the error bars represent the standard deviation estimated over the datasets. The training and testing datasets are summarized in Table \ref{tab:cubes} and described in Section \ref{sec:data}. For local optimizers, using the neural network to generate initial guesses leads to significantly lower relative fitting loss compared to using random initial guesses. Some methods achieve results comparable to PSO with $N_\text{trial}=3$, while requiring significantly fewer forward modeling runs. Among the six local optimizers tested, the gradient-free Nelder-Mead algorithm achieves the lowest relative fitting loss of $-0.01 \pm 0.09$. The SLSQP algorithm performs best among gradient-based methods, achieving a relative fitting loss of $0.00 \pm 0.09$. Additionally, the neural network can improve the performance of PSO. Using the neural network, PSO achieves better fitting results in a single run compared to PSO with $N_\text{trial} = 3$ without neural network assistance. Furthermore, no significant decrease in the number of forward modeling runs is found when using the neural network. An explanation could be that the optimizers with random initial guesses tend to get stuck in local minima, and therefore stop earlier.
\par
In Figure \ref{fig:err_nfev_2}, we demonstrate the effect of using different number of initial points generated by the neural network, with $N_\text{init}=10, 25, 50, 100$. The mean relative loss generally decreases as $N_\text{init}$ increases. However, the decrease becomes modest when $N_\text{init} \geq 50$, with improvements within 0.01, in the results based on Nelder-Mead and SLSQP.
\par
We next evaluate the generalization ability of our neural network. Figure \ref{fig:err_ds} demonstrates the mean relative loss for each dataset. As examples, we show only the results obtained using the neural network alone and the SLSQP algorithm. As described in Section \ref{sec:data}, our testing data consists of two parts. First, the data cubes labeled ATOMS-20-band3-2as, IRAS16293 A-band3-1.0as, IRAS16293 B-band3-1.0as, and Orion KL-band6-4.9as are reserved for testing and were not used during training. Secondly, the neural network is trained using only species from Mol-1980, which includes molecules detected before 1980 along with their isotopologues, while we also carry out baseline fitting for species detected between 1980 and 2010, labeled as Mol-2010. 
\par
In Figure \ref{fig:err_ds}, when using the neural network alone, the mean relative loss of the training dataset is higher than the testing dataset. However, the error can be significantly reduced by incorporating the SLSQP local optimizer in inference, resulting in a mean relative loss for the testing datasets close to zero. It is worth mentioning that the data of Orion KL-band6-4.9as are obtained from combined observations by Submillimeter Array (SMA) interferometric and IRAM 30 m single-dish telescope, while all training data were observed by ALMA, and Orion KL is not involved during the training. The mean relative loss of Orion KL-band6-4.9as is comparable to the G327 results in the training set.
\par
Furthermore, our results imply that our neural network is able to generalize for various species. Whereas we find that the mean relative loss of Mol-2010 is slightly higher than Mol-1980 for the training datasets, there is no such trend in the testing datasets. The average metrics of Mol-2010 are even smaller than Mol-1980 in the testing datasets.

\section{Applying to pixel-level fitting} \label{sec:app}
We suggest that our neural network can be efficiently applied to pixel-level fitting. As an example, we apply our method to the central hot core region of G327-band6-1.9as. The region have $100 \times 100$ pixels. We use 3 spectral windows, covering 215565 - 217273 MHz, 217572 - 219271 MHz and 229611 - 231319 MHz. This work performs fittings for each pixel independently, while our neural network based method can be applied to algorithms that take into account the correlation between the pixels. For instance, instead of fitting all pixels, \citet{2025A&A...700A..84L} introduced an active-learning framework to identify the most informative pixels for fitting, and approximated the parameter maps using the Gaussian-process model.
\par
We perform pixel-by-pixel fitting for CH$_3$OH, $v=0$, CH$_3$OCHO, $v=0$, CH$_3$OCH$_3$, $v=0$, C$_2$H$_5$CN, $v=0$, and C$_2$H$_3$CN, $v=0$. These molecules are well-established tracers of hot core chemistry \citep[e.g.,][]{2007A&A...470..639F,2024MNRAS.533.1583L} and have been previously detected in the G327 hot core \citep{2000ApJ...545..309G}.
\par
Figure \ref{fig:map_xclass} illustrates the pixel-level fitting results of CH$_3$OH, $v=0$. The first row presents the results using random initial guesses, which are quite noisy. In contrast, as shown in the second row, the results using the neural network are more accurate. In the third and fourth rows, we compare our results with \textsc{xclass} \citep{2017A&A...598A...7M}. When producing the \textsc{xclass} results, we also provide the initial parameters given by our neural network. In the central hot core, our results are consistent with \textsc{xclass}.
\par
We then compare the results obtained using the $\chi^2$ and peak matching functions in Figure \ref{fig:map_chi2_pm_CH3OH}. While the predicted column densities are consistent within $\sim 0.5$ dex, the other parameters show discrepancies. The $v_\text{offset}$ distributions exhibit similarities, while the values have offsets. Compared to the $\chi^2$ results, the results using the peak matching loss tend to underestimate the velocity width. To demonstrate the problem, we also show the spectra of two pixels in Figure \ref{fig:map_chi2_pm_CH3OH}. For clarity, we only show the spectra over a representative frequency range. It can be found that the observed spectra (black lines) include features that our spectral line model cannot reproduce, e.g. the line wings. Our model assumes only a Gaussian line profile. The $\chi^2$ and peak matching loss functions fit the observed spectra in different ways. According to \cite{2025ApJS..277...21Q}, the peak matching algorithm only compares the top part of the peaks,  while the $\chi^2$ function tends to fit the entire line shape. These may explain the discrepancies in the estimated parameters, in particular the velocity width and offset.
\par
In Figure \ref{fig:map_mols}, we present the fitting results of CH$_3$OCHO, $v=0$, CH$_3$OCH$_3$, $v=0$, C$_2$H$_5$CN, $v=0$, and C$_2$H$_3$CN, $v=0$. These results were obtained by combining the neural network predictions with the SLSQP algorithm. The peak matching loss function is employed for the fittings. No meaningful results can be obtained for these COMs using $\chi^2$ fitting due to line blending. This issue was discussed in \citet{2025ApJS..277...21Q}.
\par
Our method achieves reasonable results for these COMs. Clear velocity gradients are visible in Figure \ref{fig:map_mols}. Both excitation temperatures and column densities increase toward the center. These trends imply rotational motion and trace the physical structure of the G327 hot core. A detailed scientific explanation of these results is deferred to future studies, as this work focuses on demonstrating the fitting method.

\begin{table}
\centering
\caption{Runtime summary for the pixel-by-pixel fitting of different species.}
\label{tab:runtime}
\begin{tabular}{cccc}
    \hline
    \hline
    Name & Loss & Runtime & Runtime/pixel \\
    && (min) & (ms/pixel) \\
    \hline
    CH$_3$OH, $v=0$ & $\chi^2$  & 4.9 & 29 \\
    CH$_3$OH, $v=0$ & PM  & 17.5 & 105 \\
    CH$_3$OCHO, $v=0$ & PM  & 41.9 & 251 \\
    CH$_3$OCH3, $v=0$ & PM  & 19.5 & 117 \\
    C$_2$H$_5$CN, $v=0$ & PM  & 19.7 & 118 \\
    C$_2$H$_3$CN, $v=0$ & PM  & 14.8 & 89 \\
    \hline
\end{tabular}
\end{table}

\subsection{Performance}
Whereas the training of our neural network requires dozens of GPUs as mentioned in Section \ref{sec:training}, the inference only requires one consumer-grade GPU. The pixel-level fitting results presented above were performed using one RTX 2080 Ti card and 16 cores on a desktop. The fitting region has 10,000 pixels, with 8740 frequency channels. The execution time for pixel-by-pixel fitting of different species ranges from 4.9 minutes (29 ms/pixel) to 41.9 minutes (251 ms/pixel), as summarized in Table \ref{tab:runtime}. For methanol, the runtime using the peak matching loss is long than using the $\chi^2$ loss. The peak matching loss is less smooth than the $\chi^2$ function and, on average, results in 2.4 times forward modeling runs during fitting. Its calculation is also slower than the $\chi^2$ function.

\section{Discussion} \label{sec:discuss}
This work trains a neural network to predict the five parameters of the one-dimensional LTE radiative transfer model. The model is described in Section \ref{sec:sl_model}. This model is suitable for COMs in star-forming regions with high H$_2$ volume density \citep[e.g][]{2020A&A...641A..54C,2024A&A...686A.289H}, where the LTE assumption holds. In addition, due to degeneracies among source size, excitation temperature, and column density, our method is unsuitable for those species with only a few emission lines in the spectrum, for instance CO and HCN.
\par
We use real observational data to train the neural network, which presents both strengths and limitations. This approach enables the neural network to adapt to various observational features that are difficult to reproduce with spectral line models. However, we include only a small portion of the AMLA data, and the observational data used for training are limited by frequency coverage, beam size variations, and noise levels (see Table \ref{tab:cubes}). Incorporating more diverse observational data could reduce potential bias.
\par
Furthermore, this study considers only molecular rotational spectra in the radio band, while molecules can also be detected via vibrational transitions in the infrared band using the James Webb Space Telescope (JWST) \citep[e.g.][]{2023NatAs...7..431M,2024A&A...690A.205C,2024A&A...686A..71N}. There are two main challenges in extending our methodology to infrared observations. First, our approach relies on the efficient generation of synthetic spectra for comparison with observations, while modeling infrared transitions is complex due to contributions from both gas and ice phases \citep{2024A&A...690A.205C}. Secondly, the baseline of infrared spectra is complex\citep[e.g.][]{2023NatAs...7..431M,2024A&A...690A.205C}, whereas our peak-matching algorithm assumes a flat baseline. In such cases, local regression techniques or autoregressive moving average models \citep[e.g.][]{chatfield2019analysis,hyndman2021} could be used to correct baselines and extract spectral line features.

\section{Summary} \label{sec:summary}
In summary, this work proposes a reinforcement learning framework for spectral line fitting of interstellar molecules. The framework consists of four key components:
\begin{enumerate}
    \item An algorithm that converts the input spectrum and molecular spectroscopic data into a pair of variable-length sequences suitable for processing by neural networks.
    \item A transformer-based neural network that embeds the input sequences of spectrum and molecular spectroscopic data into a vector representation, and then decodes this vector to predict favorable fitting parameters.
    \item A training method based on policy gradients with a prioritized replay buffer.
    \item An inference method that combines the neural network with local optimization algorithms.
\end{enumerate}
\par
Compared to PSO with multiple runs, our neural network-based method achieves consistent fitting results while reducing the number of forward modeling runs by an order of magnitude.
\par
We then apply our method to pixel-level fitting of data cubes and compare the results with \textsc{xclass}. Our findings are summarized as follows:
\begin{enumerate}
    \item Using the neural network to predict the initial fitting parameters significantly improves the results.
    \item When using the $\chi^2$ function for fitting, our results are consistent with \textsc{xclass}.
    \item Results obtained using the $\chi^2$ and peak matching loss functions show different parameter estimates. Both methods fit the observed spectra differently, and we attribute these discrepancies to the fact that observations contain features that our spectral line model cannot describe.
\end{enumerate}
\par
This work presents a robust and efficient method for spectral line fitting of interstellar molecules, providing an excellent tool for analyzing data from large surveys, such as ATOMS \citep{2020MNRAS.496.2790L}, QUARKS \citep{2024RAA....24b5009L}, ALMA-IMF\citep{2022A&A...662A...8M}, and ALMAGAL \citep{2025A&A...696A.149M}.

\begin{acknowledgments}
This paper makes use of the following ALMA data: \#2012.1.00712.S, \#2015.1.01193.S, \#2016.1.00168.S, \#2017.1.00108.S, \#2018.1.01496.S, \#2019.1.00685.S, \#2021.1.01164.S, \#2022.1.01354.S, \#2024.1.00653.S. ALMA is a partnership of ESO (representing its member states), NSF (USA) and NINS (Japan), together with NRC (Canada), MOST and ASIAA (Taiwan), and KASI (Republic of Korea), in cooperation with the Republic of Chile. The Joint ALMA Observatory is operated by ESO, AUI/NRAO and NAOJ. The National Radio Astronomy Observatory is a facility of the National Science Foundation operated under cooperative agreement by Associated Universities, Inc.
\par
This work was supported by the National Natural Science Foundation of China (Grant No. 12373026), the Leading Innovation and Entrepreneurship Team of Zhejiang Province of China (Grant No. 2023R01008), the Key R\&D Program of Zhejiang, China (Grant No. 2024SSYS0012), the Young Scientists Fund of the National Natural Science Foundation of China (Grant No. 12403030), and the China Postdoctoral Science Foundation (Grant No. 2023TQ0330).

\end{acknowledgments}





%
\facilities{ALMA, SMA, IRAM:30m}

\software{\textsc{astropy} \citep{2013A&A...558A..33A,2018AJ....156..123A,2022ApJ...935..167A}, \textsc{casa} \citep{2022PASP..134k4501C}, \textsc{matplotlib} \citep{Hunter:2007}, \textsc{numpy} \citep{harris2020array}, \textsc{pandas} \citep{mckinney-proc-scipy-2010,reback2020pandas}, \textsc{scipy} \citep{2020SciPy-NMeth}, \textsc{statcont} \citep{2018A&A...609A.101S}, \textsc{xclass} \citep{2017A&A...598A...7M}.
}



\bibliography{references}{}

\begin{thebibliography}{}
\expandafter\ifx\csname natexlab\endcsname\relax\def\natexlab#1{#1}\fi
\providecommand{\url}[1]{\href{#1}{#1}}
\providecommand{\dodoi}[1]{doi:~\href{http://doi.org/#1}{\nolinkurl{#1}}}
\providecommand{\doeprint}[1]{\href{http://ascl.net/#1}{\nolinkurl{http://ascl.net/#1}}}
\providecommand{\doarXiv}[1]{\href{https://arxiv.org/abs/#1}{\nolinkurl{https://arxiv.org/abs/#1}}}

\bibitem[{J. {Alsing} {et~al.}(2019){Alsing}, {Charnock}, {Feeney}, \& {Wandelt}}]{2019MNRAS.488.4440A}
{Alsing}, J., {Charnock}, T., {Feeney}, S., \& {Wandelt}, B. 2019, \bibinfo{title}{{Fast likelihood-free cosmology with neural density estimators and active learning},} \mnras, 488, 4440, \dodoi{10.1093/mnras/stz1960}

\bibitem[{ {Astropy Collaboration} {et~al.}(2013){Astropy Collaboration}, {Robitaille}, {Tollerud}, {Greenfield}, {Droettboom}, {Bray}, {Aldcroft}, {Davis}, {Ginsburg}, {Price-Whelan}, {Kerzendorf}, {Conley}, {Crighton}, {Barbary}, {Muna}, {Ferguson}, {Grollier}, {Parikh}, {Nair}, {Unther}, {Deil}, {Woillez}, {Conseil}, {Kramer}, {Turner}, {Singer}, {Fox}, {Weaver}, {Zabalza}, {Edwards}, {Azalee Bostroem}, {Burke}, {Casey}, {Crawford}, {Dencheva}, {Ely}, {Jenness}, {Labrie}, {Lim}, {Pierfederici}, {Pontzen}, {Ptak}, {Refsdal}, {Servillat}, \& {Streicher}}]{2013A&A...558A..33A}
{Astropy Collaboration}, {Robitaille}, T.~P., {Tollerud}, E.~J., {et~al.} 2013, \bibinfo{title}{{Astropy: A community Python package for astronomy},} \aap, 558, A33, \dodoi{10.1051/0004-6361/201322068}

\bibitem[{ {Astropy Collaboration} {et~al.}(2018){Astropy Collaboration}, {Price-Whelan}, {Sip{\H{o}}cz}, {G{\"u}nther}, {Lim}, {Crawford}, {Conseil}, {Shupe}, {Craig}, {Dencheva}, {Ginsburg}, {VanderPlas}, {Bradley}, {P{\'e}rez-Su{\'a}rez}, {de Val-Borro}, {Aldcroft}, {Cruz}, {Robitaille}, {Tollerud}, {Ardelean}, {Babej}, {Bach}, {Bachetti}, {Bakanov}, {Bamford}, {Barentsen}, {Barmby}, {Baumbach}, {Berry}, {Biscani}, {Boquien}, {Bostroem}, {Bouma}, {Brammer}, {Bray}, {Breytenbach}, {Buddelmeijer}, {Burke}, {Calderone}, {Cano Rodr{\'\i}guez}, {Cara}, {Cardoso}, {Cheedella}, {Copin}, {Corrales}, {Crichton}, {D'Avella}, {Deil}, {Depagne}, {Dietrich}, {Donath}, {Droettboom}, {Earl}, {Erben}, {Fabbro}, {Ferreira}, {Finethy}, {Fox}, {Garrison}, {Gibbons}, {Goldstein}, {Gommers}, {Greco}, {Greenfield}, {Groener}, {Grollier}, {Hagen}, {Hirst}, {Homeier}, {Horton}, {Hosseinzadeh}, {Hu}, {Hunkeler}, {Ivezi{\'c}}, {Jain}, {Jenness}, {Kanarek}, {Kendrew}, {Kern}, {Kerzendorf}, {Khvalko}, {King}, {Kirkby}, {Kulkarni},
  {Kumar}, {Lee}, {Lenz}, {Littlefair}, {Ma}, {Macleod}, {Mastropietro}, {McCully}, {Montagnac}, {Morris}, {Mueller}, {Mumford}, {Muna}, {Murphy}, {Nelson}, {Nguyen}, {Ninan}, {N{\"o}the}, {Ogaz}, {Oh}, {Parejko}, {Parley}, {Pascual}, {Patil}, {Patil}, {Plunkett}, {Prochaska}, {Rastogi}, {Reddy Janga}, {Sabater}, {Sakurikar}, {Seifert}, {Sherbert}, {Sherwood-Taylor}, {Shih}, {Sick}, {Silbiger}, {Singanamalla}, {Singer}, {Sladen}, {Sooley}, {Sornarajah}, {Streicher}, {Teuben}, {Thomas}, {Tremblay}, {Turner}, {Terr{\'o}n}, {van Kerkwijk}, {de la Vega}, {Watkins}, {Weaver}, {Whitmore}, {Woillez}, {Zabalza}, \& {Astropy Contributors}}]{2018AJ....156..123A}
{Astropy Collaboration}, {Price-Whelan}, A.~M., {Sip{\H{o}}cz}, B.~M., {et~al.} 2018, \bibinfo{title}{{The Astropy Project: Building an Open-science Project and Status of the v2.0 Core Package},} \aj, 156, 123, \dodoi{10.3847/1538-3881/aabc4f}

\bibitem[{ {Astropy Collaboration} {et~al.}(2022){Astropy Collaboration}, {Price-Whelan}, {Lim}, {Earl}, {Starkman}, {Bradley}, {Shupe}, {Patil}, {Corrales}, {Brasseur}, {N{\"o}the}, {Donath}, {Tollerud}, {Morris}, {Ginsburg}, {Vaher}, {Weaver}, {Tocknell}, {Jamieson}, {van Kerkwijk}, {Robitaille}, {Merry}, {Bachetti}, {G{\"u}nther}, {Aldcroft}, {Alvarado-Montes}, {Archibald}, {B{\'o}di}, {Bapat}, {Barentsen}, {Baz{\'a}n}, {Biswas}, {Boquien}, {Burke}, {Cara}, {Cara}, {Conroy}, {Conseil}, {Craig}, {Cross}, {Cruz}, {D'Eugenio}, {Dencheva}, {Devillepoix}, {Dietrich}, {Eigenbrot}, {Erben}, {Ferreira}, {Foreman-Mackey}, {Fox}, {Freij}, {Garg}, {Geda}, {Glattly}, {Gondhalekar}, {Gordon}, {Grant}, {Greenfield}, {Groener}, {Guest}, {Gurovich}, {Handberg}, {Hart}, {Hatfield-Dodds}, {Homeier}, {Hosseinzadeh}, {Jenness}, {Jones}, {Joseph}, {Kalmbach}, {Karamehmetoglu}, {Ka{\l}uszy{\'n}ski}, {Kelley}, {Kern}, {Kerzendorf}, {Koch}, {Kulumani}, {Lee}, {Ly}, {Ma}, {MacBride}, {Maljaars}, {Muna}, {Murphy}, {Norman},
  {O'Steen}, {Oman}, {Pacifici}, {Pascual}, {Pascual-Granado}, {Patil}, {Perren}, {Pickering}, {Rastogi}, {Roulston}, {Ryan}, {Rykoff}, {Sabater}, {Sakurikar}, {Salgado}, {Sanghi}, {Saunders}, {Savchenko}, {Schwardt}, {Seifert-Eckert}, {Shih}, {Jain}, {Shukla}, {Sick}, {Simpson}, {Singanamalla}, {Singer}, {Singhal}, {Sinha}, {Sip{\H{o}}cz}, {Spitler}, {Stansby}, {Streicher}, {{\v{S}}umak}, {Swinbank}, {Taranu}, {Tewary}, {Tremblay}, {de Val-Borro}, {Van Kooten}, {Vasovi{\'c}}, {Verma}, {de Miranda Cardoso}, {Williams}, {Wilson}, {Winkel}, {Wood-Vasey}, {Xue}, {Yoachim}, {Zhang}, {Zonca}, \& {Astropy Project Contributors}}]{2022ApJ...935..167A}
{Astropy Collaboration}, {Price-Whelan}, A.~M., {Lim}, P.~L., {et~al.} 2022, \bibinfo{title}{{The Astropy Project: Sustaining and Growing a Community-oriented Open-source Project and the Latest Major Release (v5.0) of the Core Package},} \apj, 935, 167, \dodoi{10.3847/1538-4357/ac7c74}

\bibitem[{H. {Beuther} {et~al.}(2020){Beuther}, {Soler}, {Linz}, {Henning}, {Gieser}, {Kuiper}, {Vlemmings}, {Hennebelle}, {Feng}, {Smith}, \& {Ahmadi}}]{2020ApJ...904..168B}
{Beuther}, H., {Soler}, J.~D., {Linz}, H., {et~al.} 2020, \bibinfo{title}{{Gravity and Rotation Drag the Magnetic Field in High-mass Star Formation},} \apj, 904, 168, \dodoi{10.3847/1538-4357/abc019}

\bibitem[{U. {Bhardwaj} {et~al.}(2023){Bhardwaj}, {Alvey}, {Miller}, {Nissanke}, \& {Weniger}}]{2023PhRvD.108d2004B}
{Bhardwaj}, U., {Alvey}, J., {Miller}, B.~K., {Nissanke}, S., \& {Weniger}, C. 2023, \bibinfo{title}{{Sequential simulation-based inference for gravitational wave signals},} \prd, 108, 042004, \dodoi{10.1103/PhysRevD.108.042004}

\bibitem[{L. Breiman(2001)Breiman}]{breiman2001random}
Breiman, L. 2001, \bibinfo{title}{Random forests,} Machine learning, 45, 5

\bibitem[{R.~H. Byrd {et~al.}(1995)Byrd, Lu, Nocedal, \& Zhu}]{byrd1995limited}
Byrd, R.~H., Lu, P., Nocedal, J., \& Zhu, C. 1995, \bibinfo{title}{A limited memory algorithm for bound constrained optimization,} SIAM Journal on scientific computing, 16, 1190

\bibitem[{J. {Carpenter} {et~al.}(2020){Carpenter}, {Iono}, {Kemper}, \& {Wootten}}]{2020arXiv200111076C}
{Carpenter}, J., {Iono}, D., {Kemper}, F., \& {Wootten}, A. 2020, \bibinfo{title}{{The ALMA Development Program: Roadmap to 2030},} arXiv e-prints, arXiv:2001.11076, \dodoi{10.48550/arXiv.2001.11076}

\bibitem[{ {CASA Team} {et~al.}(2022){CASA Team}, {Bean}, {Bhatnagar}, {Castro}, {Donovan Meyer}, {Emonts}, {Garcia}, {Garwood}, {Golap}, {Gonzalez Villalba}, {Harris}, {Hayashi}, {Hoskins}, {Hsieh}, {Jagannathan}, {Kawasaki}, {Keimpema}, {Kettenis}, {Lopez}, {Marvil}, {Masters}, {McNichols}, {Mehringer}, {Miel}, {Moellenbrock}, {Montesino}, {Nakazato}, {Ott}, {Petry}, {Pokorny}, {Raba}, {Rau}, {Schiebel}, {Schweighart}, {Sekhar}, {Shimada}, {Small}, {Steeb}, {Sugimoto}, {Suoranta}, {Tsutsumi}, {van Bemmel}, {Verkouter}, {Wells}, {Xiong}, {Szomoru}, {Griffith}, {Glendenning}, \& {Kern}}]{2022PASP..134k4501C}
{CASA Team}, {Bean}, B., {Bhatnagar}, S., {et~al.} 2022, \bibinfo{title}{{CASA, the Common Astronomy Software Applications for Radio Astronomy},} \pasp, 134, 114501, \dodoi{10.1088/1538-3873/ac9642}

\bibitem[{C. Chatfield \& H. Xing(2019)Chatfield \& Xing}]{chatfield2019analysis}
Chatfield, C., \& Xing, H. 2019, The analysis of time series: an introduction with R (Chapman and hall/CRC)

\bibitem[{Y. {Chen} {et~al.}(2024){Chen}, {Rocha}, {van Dishoeck}, {van Gelder}, {Nazari}, {Slavicinska}, {Francis}, {Tabone}, {Ressler}, {Klaassen}, {Beuther}, {Boogert}, {Gieser}, {Kavanagh}, {Perotti}, {Le Gouellec}, {Majumdar}, {G{\"u}del}, \& {Henning}}]{2024A&A...690A.205C}
{Chen}, Y., {Rocha}, W.~R.~M., {van Dishoeck}, E.~F., {et~al.} 2024, \bibinfo{title}{{JOYS+: The link between the ice and gas of complex organic molecules: Comparing JWST and ALMA data of two low-mass protostars},} \aap, 690, A205, \dodoi{10.1051/0004-6361/202450706}

\bibitem[{A. {Coletta} {et~al.}(2020){Coletta}, {Fontani}, {Rivilla}, {Mininni}, {Colzi}, {S{\'a}nchez-Monge}, \& {Beltr{\'a}n}}]{2020A&A...641A..54C}
{Coletta}, A., {Fontani}, F., {Rivilla}, V.~M., {et~al.} 2020, \bibinfo{title}{{Evolutionary study of complex organic molecules in high-mass star-forming regions},} \aap, 641, A54, \dodoi{10.1051/0004-6361/202038212}

\bibitem[{J. Devlin {et~al.}(2019)Devlin, Chang, Lee, \& Toutanova}]{devlin-etal-2019-bert}
Devlin, J., Chang, M.-W., Lee, K., \& Toutanova, K. 2019, in Proceedings of the 2019 Conference of the North {A}merican Chapter of the Association for Computational Linguistics: Human Language Technologies, Volume 1 (Long and Short Papers), ed. J.~Burstein, C.~Doran, \& T.~Solorio (Minneapolis, Minnesota: Association for Computational Linguistics), 4171--4186, \dodoi{10.18653/v1/N19-1423}

\bibitem[{A. Dosovitskiy {et~al.}(2021)Dosovitskiy, Beyer, Kolesnikov, Weissenborn, Zhai, Unterthiner, Dehghani, Minderer, Heigold, Gelly, Uszkoreit, \& Houlsby}]{dosovitskiy2021an}
Dosovitskiy, A., Beyer, L., Kolesnikov, A., {et~al.} 2021, in International Conference on Learning Representations.
\newblock \url{https://openreview.net/forum?id=YicbFdNTTy}

\bibitem[{S.~A. {Dzib} {et~al.}(2018){Dzib}, {Ortiz-Le{\'o}n}, {Hern{\'a}ndez-G{\'o}mez}, {Loinard}, {Mioduszewski}, {Claussen}, {Menten}, {Caux}, \& {Sanna}}]{2018A&A...614A..20D}
{Dzib}, S.~A., {Ortiz-Le{\'o}n}, G.~N., {Hern{\'a}ndez-G{\'o}mez}, A., {et~al.} 2018, \bibinfo{title}{{A revised distance to IRAS 16293-2422 from VLBA astrometry of associated water masers},} \aap, 614, A20, \dodoi{10.1051/0004-6361/201732093}

\bibitem[{L. {Einig} {et~al.}(2024){Einig}, {Palud}, {Roueff}, {Pety}, {Bron}, {Le Petit}, {Gerin}, {Chanussot}, {Chainais}, {Thouvenin}, {Languignon}, {Be{\v{s}}li{\'c}}, {Coud{\'e}}, {Mazurek}, {Orkisz}, {Santa-Maria}, {S{\'e}gal}, {Zakardjian}, {Bardeau}, {Demyk}, {de Souza Magalh{\~a}es}, {Goicoechea}, {Gratier}, {Guzm{\'a}n}, {Hughes}, {Levrier}, {Le Bourlot}, {Lis}, {Liszt}, {Peretto}, {Roueff}, \& {Sievers}}]{2024A&A...691A.109E}
{Einig}, L., {Palud}, P., {Roueff}, A., {et~al.} 2024, \bibinfo{title}{{Quantifying the informativity of emission lines to infer physical conditions in giant molecular clouds: I. Application to model predictions},} \aap, 691, A109, \dodoi{10.1051/0004-6361/202451588}

\bibitem[{S.~J. {El-Abd} {et~al.}(2024){El-Abd}, {Brogan}, {Hunter}, {Lee}, {Loomis}, \& {McGuire}}]{2024ApJ...965...14E}
{El-Abd}, S.~J., {Brogan}, C.~L., {Hunter}, T.~R., {et~al.} 2024, \bibinfo{title}{{An Automated Chemical Exploration of NGC 6334I at 340 au Resolution},} \apj, 965, 14, \dodoi{10.3847/1538-4357/ad283f}

\bibitem[{C.~P. {Endres} {et~al.}(2016){Endres}, {Schlemmer}, {Schilke}, {Stutzki}, \& {M{\"u}ller}}]{2016JMoSp.327...95E}
{Endres}, C.~P., {Schlemmer}, S., {Schilke}, P., {Stutzki}, J., \& {M{\"u}ller}, H. S.~P. 2016, \bibinfo{title}{{The Cologne Database for Molecular Spectroscopy, CDMS, in the Virtual Atomic and Molecular Data Centre, VAMDC},} Journal of Molecular Spectroscopy, 327, 95, \dodoi{10.1016/j.jms.2016.03.005}

\bibitem[{S. {Feng} {et~al.}(2015){Feng}, {Beuther}, {Henning}, {Semenov}, {Palau}, \& {Mills}}]{2015A&A...581A..71F}
{Feng}, S., {Beuther}, H., {Henning}, T., {et~al.} 2015, \bibinfo{title}{{Resolving the chemical substructure of Orion-KL},} \aap, 581, A71, \dodoi{10.1051/0004-6361/201322725}

\bibitem[{F. {Fontani} {et~al.}(2007){Fontani}, {Pascucci}, {Caselli}, {Wyrowski}, {Cesaroni}, \& {Walmsley}}]{2007A&A...470..639F}
{Fontani}, F., {Pascucci}, I., {Caselli}, P., {et~al.} 2007, \bibinfo{title}{{Comparative study of complex N- and O-bearing molecules in hot molecular cores},} \aap, 470, 639, \dodoi{10.1051/0004-6361:20077485}

\bibitem[{F. Gao \& L. Han(2012)Gao \& Han}]{gao2012implementing}
Gao, F., \& Han, L. 2012, \bibinfo{title}{Implementing the Nelder-Mead simplex algorithm with adaptive parameters,} Computational Optimization and Applications, 51, 259

\bibitem[{E. {Gibb} {et~al.}(2000){Gibb}, {Nummelin}, {Irvine}, {Whittet}, \& {Bergman}}]{2000ApJ...545..309G}
{Gibb}, E., {Nummelin}, A., {Irvine}, W.~M., {Whittet}, D.~C.~B., \& {Bergman}, P. 2000, \bibinfo{title}{{Chemistry of the Organic-Rich Hot Core G327.3-0.6},} \apj, 545, 309, \dodoi{10.1086/317805}

\bibitem[{A. {Ginsburg} {et~al.}(2022){Ginsburg}, {Sokolov}, {de Val-Borro}, {Rosolowsky}, {Pineda}, {Sip{\H{o}}cz}, \& {Henshaw}}]{2022AJ....163..291G}
{Ginsburg}, A., {Sokolov}, V., {de Val-Borro}, M., {et~al.} 2022, \bibinfo{title}{{Pyspeckit: A Spectroscopic Analysis and Plotting Package},} \aj, 163, 291, \dodoi{10.3847/1538-3881/ac695a}

\bibitem[{P.~F. {Goldsmith} \& W.~D. {Langer}(1999){Goldsmith} \& {Langer}}]{1999ApJ...517..209G}
{Goldsmith}, P.~F., \& {Langer}, W.~D. 1999, \bibinfo{title}{{Population Diagram Analysis of Molecular Line Emission},} \apj, 517, 209, \dodoi{10.1086/307195}

\bibitem[{I. Goodfellow {et~al.}(2016)Goodfellow, Bengio, \& Courville}]{Goodfellow-et-al-2016}
Goodfellow, I., Bengio, Y., \& Courville, A. 2016, Deep Learning (MIT Press)

\bibitem[{T. {Grassi} {et~al.}(2025){Grassi}, {Padovani}, {Galli}, {Vaytet}, {Jensen}, {Redaelli}, {Spezzano}, {Bovino}, \& {Caselli}}]{2025A&A...702A..71G}
{Grassi}, T., {Padovani}, M., {Galli}, D., {et~al.} 2025, \bibinfo{title}{{Mapping synthetic observations to pre-stellar core models: An interpretable machine learning approach},} \aap, 702, A71, \dodoi{10.1051/0004-6361/202453266}

\bibitem[{C.~R. Harris {et~al.}(2020)Harris, Millman, van~der Walt, Gommers, Virtanen, Cournapeau, Wieser, Taylor, Berg, Smith, Kern, Picus, Hoyer, van Kerkwijk, Brett, Haldane, del R{\'{i}}o, Wiebe, Peterson, G{\'{e}}rard-Marchant, Sheppard, Reddy, Weckesser, Abbasi, Gohlke, \& Oliphant}]{harris2020array}
Harris, C.~R., Millman, K.~J., van~der Walt, S.~J., {et~al.} 2020, \bibinfo{title}{Array programming with {NumPy},} Nature, 585, 357, \dodoi{10.1038/s41586-020-2649-2}

\bibitem[{J. {Heyl} {et~al.}(2023){Heyl}, {Butterworth}, \& {Viti}}]{2023MNRAS.526..404H}
{Heyl}, J., {Butterworth}, J., \& {Viti}, S. 2023, \bibinfo{title}{{Understanding molecular abundances in star-forming regions using interpretable machine learning},} \mnras, 526, 404, \dodoi{10.1093/mnras/stad2814}

\bibitem[{D. Horgan {et~al.}(2018)Horgan, Quan, Budden, Barth-Maron, Hessel, van Hasselt, \& Silver}]{horgan2018distributed}
Horgan, D., Quan, J., Budden, D., {et~al.} 2018, in International Conference on Learning Representations.
\newblock \url{https://openreview.net/forum?id=H1Dy---0Z}

\bibitem[{T.-H. {Hsieh} {et~al.}(2024){Hsieh}, {Pineda}, {Segura-Cox}, {Caselli}, {Valdivia-Mena}, {Gieser}, {Maureira}, {Lopez-Sepulcre}, {Bouscasse}, {Neri}, {M{\"o}ller}, {Dutrey}, {Fuente}, {Semenov}, {Chapillon}, {Cunningham}, {Henning}, {Pi{\'e}tu}, {Jimenez-Serra}, {Marino}, \& {Ceccarelli}}]{2024A&A...686A.289H}
{Hsieh}, T.-H., {Pineda}, J.~E., {Segura-Cox}, D.~M., {et~al.} 2024, \bibinfo{title}{{PRODIGE - envelope to disk with NOEMA. III. The origin of complex organic molecule emission in SVS13A},} \aap, 686, A289, \dodoi{10.1051/0004-6361/202449417}

\bibitem[{J.~D. Hunter(2007)Hunter}]{Hunter:2007}
Hunter, J.~D. 2007, \bibinfo{title}{Matplotlib: A 2D graphics environment,} Computing in Science \& Engineering, 9, 90, \dodoi{10.1109/MCSE.2007.55}

\bibitem[{R. Hyndman \& G. Athanasopoulos(2021)Hyndman \& Athanasopoulos}]{hyndman2021}
Hyndman, R., \& Athanasopoulos, G. 2021, Forecasting: principles and practice, 3rd edn. (Melbourne, Australia: OTexts).
\newblock \url{https://OTexts.com/fpp3}

\bibitem[{J.~K. {J{\o}rgensen} {et~al.}(2020){J{\o}rgensen}, {Belloche}, \& {Garrod}}]{2020ARA&A..58..727J}
{J{\o}rgensen}, J.~K., {Belloche}, A., \& {Garrod}, R.~T. 2020, \bibinfo{title}{{Astrochemistry During the Formation of Stars},} \araa, 58, 727, \dodoi{10.1146/annurev-astro-032620-021927}

\bibitem[{N. {Kessler} {et~al.}(2025){Kessler}, {Csengeri}, {Cornu}, {Bontemps}, \& {Bouscasse}}]{2025arXiv251009119K}
{Kessler}, N., {Csengeri}, T., {Cornu}, D., {Bontemps}, S., \& {Bouscasse}, L. 2025, \bibinfo{title}{{Identification of molecular line emission using Convolutional Neural Networks},} arXiv e-prints, arXiv:2510.09119, \dodoi{10.48550/arXiv.2510.09119}

\bibitem[{D. Kraft(1988)Kraft}]{kraft1988software}
Kraft, D. 1988, A Software Package for Sequential Quadratic Programming, Deutsche Forschungs- und Versuchsanstalt f{\"u}r Luft- und Raumfahrt K{\"o}ln: Forschungsbericht (Wiss. Berichtswesen d. DFVLR).
\newblock \url{https://books.google.com.hk/books?id=4rKaGwAACAAJ}

\bibitem[{K.~L.~K. {Lee} {et~al.}(2021){Lee}, {Patterson}, {Burkhardt}, {Vankayalapati}, {McCarthy}, \& {McGuire}}]{2021ApJ...917L...6L}
{Lee}, K. L.~K., {Patterson}, J., {Burkhardt}, A.~M., {et~al.} 2021, \bibinfo{title}{{Machine Learning of Interstellar Chemical Inventories},} \apjl, 917, L6, \dodoi{10.3847/2041-8213/ac194b}

\bibitem[{S. {Leurini} {et~al.}(2017){Leurini}, {Herpin}, {van der Tak}, {Wyrowski}, {Herczeg}, \& {van Dishoeck}}]{2017A&A...602A..70L}
{Leurini}, S., {Herpin}, F., {van der Tak}, F., {et~al.} 2017, \bibinfo{title}{{Distribution of water in the G327.3-0.6 massive star-forming region},} \aap, 602, A70, \dodoi{10.1051/0004-6361/201730387}

\bibitem[{C. {Li} {et~al.}(2024){Li}, {Qin}, {Liu}, {Liu}, {Tang}, {Liu}, {Chen}, {Li}, {Xu}, {Zhang}, {Liu}, {Shi}, \& {Wu}}]{2024MNRAS.533.1583L}
{Li}, C., {Qin}, S.-L., {Liu}, T., {et~al.} 2024, \bibinfo{title}{{Correlations of methyl formate (CH$_{3}$OCHO), dimethyl ether (CH$_{3}$OCH$_{3}$), and ketene (H$_{2}$CCO) in high-mass star-forming regions},} \mnras, 533, 1583, \dodoi{10.1093/mnras/stae1934}

\bibitem[{Y. {Lin} {et~al.}(2025){Lin}, {Adachi}, {Spezzano}, {Edenhofer}, {Eberle}, {Osborne}, \& {Caselli}}]{2025A&A...700A..84L}
{Lin}, Y., {Adachi}, M., {Spezzano}, S., {et~al.} 2025, \bibinfo{title}{{BASIL: Fast broadband line-rich spectral-cube fitting and image visualization via Bayesian quadrature},} \aap, 700, A84, \dodoi{10.1051/0004-6361/202452828}

\bibitem[{T. {Liu} {et~al.}(2020){Liu}, {Evans}, {Kim}, {Goldsmith}, {Liu}, {Zhang}, {Tatematsu}, {Wang}, {Juvela}, {Bronfman}, {Cunningham}, {Garay}, {Hirota}, {Lee}, {Kang}, {Li}, {Li}, {Mardones}, {Qin}, {Ristorcelli}, {Tej}, {Toth}, {Wu}, {Wu}, {Yi}, {Yun}, {Liu}, {Peng}, {Li}, {Li}, {Lee}, {Shen}, {Baug}, {Wang}, {Zhang}, {Issac}, {Zhu}, {Luo}, {Soam}, {Liu}, {Xu}, {Wang}, {Zhang}, {Ren}, \& {Zhang}}]{2020MNRAS.496.2790L}
{Liu}, T., {Evans}, N.~J., {Kim}, K.-T., {et~al.} 2020, \bibinfo{title}{{ATOMS: ALMA Three-millimeter Observations of Massive Star-forming regions - I. Survey description and a first look at G9.62+0.19},} \mnras, 496, 2790, \dodoi{10.1093/mnras/staa1577}

\bibitem[{X. {Liu} {et~al.}(2024){Liu}, {Liu}, {Zhu}, {Garay}, {Liu}, {Goldsmith}, {Evans}, {Kim}, {Liu}, {Xu}, {Lu}, {Tej}, {Mai}, {Bronfman}, {Li}, {Mardones}, {Stutz}, {Tatematsu}, {Wang}, {Zhang}, {Qin}, {Zhou}, {Luo}, {Zhang}, {Cheng}, {He}, {Gu}, {Li}, {Zhang}, {Zhang}, {Saha}, {Dewangan}, {Sanhueza}, \& {Shen}}]{2024RAA....24b5009L}
{Liu}, X., {Liu}, T., {Zhu}, L., {et~al.} 2024, \bibinfo{title}{{The ALMA-QUARKS Survey. I. Survey Description and Data Reduction},} Research in Astronomy and Astrophysics, 24, 025009, \dodoi{10.1088/1674-4527/ad0d5c}

\bibitem[{I. Loshchilov \& F. Hutter(2017)Loshchilov \& Hutter}]{loshchilov2017decoupled}
Loshchilov, I., \& Hutter, F. 2017, \bibinfo{title}{Decoupled weight decay regularization,} arXiv preprint arXiv:1711.05101

\bibitem[{S. {Maret} {et~al.}(2011){Maret}, {Hily-Blant}, {Pety}, {Bardeau}, \& {Reynier}}]{2011A&A...526A..47M}
{Maret}, S., {Hily-Blant}, P., {Pety}, J., {Bardeau}, S., \& {Reynier}, E. 2011, \bibinfo{title}{{Weeds: a CLASS extension for the analysis of millimeter and sub-millimeter spectral surveys},} \aap, 526, A47, \dodoi{10.1051/0004-6361/201015487}

\bibitem[{S. {Mart{\'\i}n} {et~al.}(2019){Mart{\'\i}n}, {Mart{\'\i}n-Pintado}, {Blanco-S{\'a}nchez}, {Rivilla}, {Rodr{\'\i}guez-Franco}, \& {Rico-Villas}}]{2019A&A...631A.159M}
{Mart{\'\i}n}, S., {Mart{\'\i}n-Pintado}, J., {Blanco-S{\'a}nchez}, C., {et~al.} 2019, \bibinfo{title}{{Spectral Line Identification and Modelling (SLIM) in the MAdrid Data CUBe Analysis (MADCUBA) package. Interactive software for data cube analysis},} \aap, 631, A159, \dodoi{10.1051/0004-6361/201936144}

\bibitem[{M.~K. {McClure} {et~al.}(2023){McClure}, {Rocha}, {Pontoppidan}, {Crouzet}, {Chu}, {Dartois}, {Lamberts}, {Noble}, {Pendleton}, {Perotti}, {Qasim}, {Rachid}, {Smith}, {Sun}, {Beck}, {Boogert}, {Brown}, {Caselli}, {Charnley}, {Cuppen}, {Dickinson}, {Drozdovskaya}, {Egami}, {Erkal}, {Fraser}, {Garrod}, {Harsono}, {Ioppolo}, {Jim{\'e}nez-Serra}, {Jin}, {J{\o}rgensen}, {Kristensen}, {Lis}, {McCoustra}, {McGuire}, {Melnick}, {{\"O}berg}, {Palumbo}, {Shimonishi}, {Sturm}, {van Dishoeck}, \& {Linnartz}}]{2023NatAs...7..431M}
{McClure}, M.~K., {Rocha}, W.~R.~M., {Pontoppidan}, K.~M., {et~al.} 2023, \bibinfo{title}{{An Ice Age JWST inventory of dense molecular cloud ices},} Nature Astronomy, 7, 431, \dodoi{10.1038/s41550-022-01875-w}

\bibitem[{B.~A. {McGuire}(2021){McGuire}}]{2021zndo...5046939M}
{McGuire}, B.~A. 2021, \bibinfo{title}{{bmcguir2/astromol: 2021 Census Update Submission},}, v2021.0.0 Zenodo, \dodoi{10.5281/zenodo.5046939}

\bibitem[{B.~A. {McGuire}(2022){McGuire}}]{2022ApJS..259...30M}
{McGuire}, B.~A. 2022, \bibinfo{title}{{2021 Census of Interstellar, Circumstellar, Extragalactic, Protoplanetary Disk, and Exoplanetary Molecules},} \apjs, 259, 30, \dodoi{10.3847/1538-4365/ac2a48}

\bibitem[{E. {Mendoza} {et~al.}(2025){Mendoza}, {Dall'Olio}, {Coelho}, {Peregr{\'\i}n}, {L{\'o}pez-Dom{\'\i}nguez}, {van der Tak}, \& {Carvajal}}]{2025A&A...698A.286M}
{Mendoza}, E., {Dall'Olio}, P., {Coelho}, L.~S., {et~al.} 2025, \bibinfo{title}{{Unmasking the physical information inherent to interstellar spectral line profiles with machine learning: I. Application of LTE to HCN and HNC transitions},} \aap, 698, A286, \dodoi{10.1051/0004-6361/202452397}

\bibitem[{V. Mnih {et~al.}(2015)Mnih, Kavukcuoglu, Silver, Rusu, Veness, Bellemare, Graves, Riedmiller, Fidjeland, Ostrovski, {et~al.}}]{mnih2015human}
Mnih, V., Kavukcuoglu, K., Silver, D., {et~al.} 2015, \bibinfo{title}{Human-level control through deep reinforcement learning,} nature, 518, 529

\bibitem[{S. {Molinari} {et~al.}(2025){Molinari}, {Schilke}, {Battersby}, {Ho}, {S{\'a}nchez-Monge}, {Traficante}, {Jones}, {Beltr{\'a}n}, {Beuther}, {Fuller}, {Zhang}, {Klessen}, {Walch}, {Tang}, {Benedettini}, {Elia}, {Coletta}, {Mininni}, {Schisano}, {Avison}, {Law}, {Nucara}, {Soler}, {Stroud}, {Wallace}, {Wells}, {Ahmadi}, {Brogan}, {Hunter}, {Liu}, {Pezzuto}, {Su}, {Zimmermann}, {Zhang}, {Wyrowski}, {De Angelis}, {Liu}, {Clarke}, {Fontani}, {Klaassen}, {Koch}, {Johnston}, {Lebreuilly}, {Liu}, {Lumsden}, {Moeller}, {Moscadelli}, {Kuiper}, {Lis}, {Peretto}, {Pfalzner}, {Rigby}, {Sanhueza}, {Rygl}, {van der Tak}, {Zinnecker}, {Amaral}, {Bally}, {Bronfman}, {Cesaroni}, {Goh}, {Hoare}, {Hatchfield}, {Hennebelle}, {Henning}, {Kim}, {Kim}, {Maud}, {Merello}, {Nakamura}, {Plume}, {Qin}, {Svoboda}, {Testi}, {Veena}, \& {Walker}}]{2025A&A...696A.149M}
{Molinari}, S., {Schilke}, P., {Battersby}, C., {et~al.} 2025, \bibinfo{title}{{ALMAGAL: I. The ALMA evolutionary study of high-mass protocluster formation in the Galaxy: Presentation of the survey and early results},} \aap, 696, A149, \dodoi{10.1051/0004-6361/202452702}

\bibitem[{T. {M{\"o}ller} {et~al.}(2017){M{\"o}ller}, {Endres}, \& {Schilke}}]{2017A&A...598A...7M}
{M{\"o}ller}, T., {Endres}, C., \& {Schilke}, P. 2017, \bibinfo{title}{{eXtended CASA Line Analysis Software Suite (XCLASS)},} \aap, 598, A7, \dodoi{10.1051/0004-6361/201527203}

\bibitem[{F. {Motte} {et~al.}(2022){Motte}, {Bontemps}, {Csengeri}, {Pouteau}, {Louvet}, {Stutz}, {Cunningham}, {L{\'o}pez-Sepulcre}, {Brouillet}, {Galv{\'a}n-Madrid}, {Ginsburg}, {Maud}, {Men'shchikov}, {Nakamura}, {Nony}, {Sanhueza}, {{\'A}lvarez-Guti{\'e}rrez}, {Armante}, {Baug}, {Bonfand}, {Busquet}, {Chapillon}, {D{\'\i}az-Gonz{\'a}lez}, {Fern{\'a}ndez-L{\'o}pez}, {Guzm{\'a}n}, {Herpin}, {Liu}, {Olguin}, {Towner}, {Bally}, {Battersby}, {Braine}, {Bronfman}, {Chen}, {Dell'Ova}, {Di Francesco}, {Gonz{\'a}lez}, {Gusdorf}, {Hennebelle}, {Izumi}, {Joncour}, {Lee}, {Lefloch}, {Lesaffre}, {Lu}, {Menten}, {Mignon-Risse}, {Molet}, {Moraux}, {Mundy}, {Nguyen Luong}, {Reyes}, {Reyes Reyes}, {Robitaille}, {Rosolowsky}, {Sandoval-Garrido}, {Schuller}, {Svoboda}, {Tatematsu}, {Thomasson}, {Walker}, {Wu}, {Whitworth}, \& {Wyrowski}}]{2022A&A...662A...8M}
{Motte}, F., {Bontemps}, S., {Csengeri}, T., {et~al.} 2022, \bibinfo{title}{{ALMA-IMF. I. Investigating the origin of stellar masses: Introduction to the Large Program and first results},} \aap, 662, A8, \dodoi{10.1051/0004-6361/202141677}

\bibitem[{H.~S.~P. {M{\"u}ller} {et~al.}(2005){M{\"u}ller}, {Schl{\"o}der}, {Stutzki}, \& {Winnewisser}}]{2005JMoSt.742..215M}
{M{\"u}ller}, H. S.~P., {Schl{\"o}der}, F., {Stutzki}, J., \& {Winnewisser}, G. 2005, \bibinfo{title}{{The Cologne Database for Molecular Spectroscopy, CDMS: a useful tool for astronomers and spectroscopists},} Journal of Molecular Structure, 742, 215, \dodoi{10.1016/j.molstruc.2005.01.027}

\bibitem[{H.~S.~P. {M{\"u}ller} {et~al.}(2001){M{\"u}ller}, {Thorwirth}, {Roth}, \& {Winnewisser}}]{2001A&A...370L..49M}
{M{\"u}ller}, H.~S.~P., {Thorwirth}, S., {Roth}, D.~A., \& {Winnewisser}, G. 2001, \bibinfo{title}{{The Cologne Database for Molecular Spectroscopy, CDMS},} \aap, 370, L49, \dodoi{10.1051/0004-6361:20010367}

\bibitem[{T. M\"{u}ller {et~al.}(2019)M\"{u}ller, Mcwilliams, Rousselle, Gross, \& Nov\'{a}k}]{mueller18neural-v2}
M\"{u}ller, T., Mcwilliams, B., Rousselle, F., Gross, M., \& Nov\'{a}k, J. 2019, \bibinfo{title}{Neural Importance Sampling,} ACM Trans. Graph., 38, \dodoi{10.1145/3341156}

\bibitem[{S.~G. Nash(1984)Nash}]{nash1984newton}
Nash, S.~G. 1984, \bibinfo{title}{Newton-type minimization via the Lanczos method,} SIAM Journal on Numerical Analysis, 21, 770

\bibitem[{P. {Nazari} {et~al.}(2024{\natexlab{a}}){Nazari}, {Cheung}, {Asensio}, {Murillo}, {van Dishoeck}, {J{\o}rgensen}, {Bourke}, {Chuang}, {Drozdovskaya}, {Fedoseev}, {Garrod}, {Ioppolo}, {Linnartz}, {McGuire}, {M{\"u}ller}, {Qasim}, \& {Wampfler}}]{2024A&A...686A..59N}
{Nazari}, P., {Cheung}, J.~S.~Y., {Asensio}, J.~F., {et~al.} 2024{\natexlab{a}}, \bibinfo{title}{{A deep search for large complex organic species toward IRAS16293-2422 B at 3 mm with ALMA},} \aap, 686, A59, \dodoi{10.1051/0004-6361/202347832}

\bibitem[{P. {Nazari} {et~al.}(2024{\natexlab{b}}){Nazari}, {Rocha}, {Rubinstein}, {Slavicinska}, {Rachid}, {van Dishoeck}, {Megeath}, {Gutermuth}, {Tyagi}, {Brunken}, {Narang}, {Manoj}, {Watson}, {Evans}, {Federman}, {Muzerolle Page}, {Anglada}, {Beuther}, {Klaassen}, {Looney}, {Osorio}, {Stanke}, \& {Yang}}]{2024A&A...686A..71N}
{Nazari}, P., {Rocha}, W.~R.~M., {Rubinstein}, A.~E., {et~al.} 2024{\natexlab{b}}, \bibinfo{title}{{Hunting for complex cyanides in protostellar ices with the JWST. A tentative detection of CH$_{3}$CN and C$_{2}$H$_{5}$CN},} \aap, 686, A71, \dodoi{10.1051/0004-6361/202348695}

\bibitem[{T. Nguyen {et~al.}(2019)Nguyen, Nguyen, Nguyen, \& Nguyen}]{nguyen2019efficient}
Nguyen, T., Nguyen, T., Nguyen, B.~M., \& Nguyen, G. 2019, \bibinfo{title}{Efficient time-series forecasting using neural network and opposition-based coral reefs optimization,} International Journal of Computational Intelligence Systems, 12, 1144

\bibitem[{T. pandas~development team(2020)pandas~development team}]{reback2020pandas}
pandas~development team, T. 2020, \bibinfo{title}{pandas-dev/pandas: Pandas,}, latest Zenodo, \dodoi{10.5281/zenodo.3509134}

\bibitem[{G. Papamakarios {et~al.}(2021)Papamakarios, Nalisnick, Rezende, Mohamed, \& Lakshminarayanan}]{papamakarios2021normalizing}
Papamakarios, G., Nalisnick, E., Rezende, D.~J., Mohamed, S., \& Lakshminarayanan, B. 2021, \bibinfo{title}{Normalizing flows for probabilistic modeling and inference,} Journal of Machine Learning Research, 22, 1

\bibitem[{G. Papamakarios {et~al.}(2017)Papamakarios, Pavlakou, \& Murray}]{NIPS2017_6c1da886}
Papamakarios, G., Pavlakou, T., \& Murray, I. 2017, in Advances in Neural Information Processing Systems, ed. I.~Guyon, U.~V. Luxburg, S.~Bengio, H.~Wallach, R.~Fergus, S.~Vishwanathan, \& R.~Garnett, Vol.~30 (Curran Associates, Inc.).
\newblock \url{https://proceedings.neurips.cc/paper_files/paper/2017/file/6c1da886822c67822bcf3679d04369fa-Paper.pdf}

\bibitem[{L. {Parker} {et~al.}(2024){Parker}, {Lanusse}, {Golkar}, {Sarra}, {Cranmer}, {Bietti}, {Eickenberg}, {Krawezik}, {McCabe}, {Morel}, {Ohana}, {Pettee}, {R{\'e}galdo-Saint Blancard}, {Cho}, {Ho}, \& {Polymathic AI Collaboration}}]{2024MNRAS.531.4990P}
{Parker}, L., {Lanusse}, F., {Golkar}, S., {et~al.} 2024, \bibinfo{title}{{AstroCLIP: a cross-modal foundation model for galaxies},} \mnras, 531, 4990, \dodoi{10.1093/mnras/stae1450}

\bibitem[{M. Patacchiola {et~al.}(2024)Patacchiola, Shysheya, Hofmann, \& Turner}]{patacchiola2024transformer}
Patacchiola, M., Shysheya, A., Hofmann, K., \& Turner, R.~E. 2024, in ICML 2024 Workshop on Structured Probabilistic Inference {\&} Generative Modeling.
\newblock \url{https://openreview.net/forum?id=mJYllFw85A}

\bibitem[{H.~M. {Pickett} {et~al.}(1998){Pickett}, {Poynter}, {Cohen}, {Delitsky}, {Pearson}, \& {M{\"u}ller}}]{1998JQSRT..60..883P}
{Pickett}, H.~M., {Poynter}, R.~L., {Cohen}, E.~A., {et~al.} 1998, \bibinfo{title}{{Submillimeter, millimeter and microwave spectral line catalog.},} \jqsrt, 60, 883, \dodoi{10.1016/S0022-4073(98)00091-0}

\bibitem[{M.~J. Powell(1994)Powell}]{powell1994direct}
Powell, M.~J. 1994, in Advances in optimization and numerical analysis (Springer), 51--67

\bibitem[{Y. {Qiu} {et~al.}(2025){Qiu}, {Zhang}, {M{\"o}ller}, {Jiang}, {Song}, {Chen}, \& {Quan}}]{2025ApJS..277...21Q}
{Qiu}, Y., {Zhang}, T., {M{\"o}ller}, T., {et~al.} 2025, \bibinfo{title}{{Spectuner: A Framework for Automated Line Identification of Interstellar Molecules},} \apjs, 277, 21, \dodoi{10.3847/1538-4365/adaeba}

\bibitem[{T.~M. Ragonneau(2022)Ragonneau}]{rago_thesis}
Ragonneau, T.~M. 2022, PhD thesis, Department of Applied Mathematics, The Hong Kong Polytechnic University, Hong Kong, China.
\newblock \url{https://theses.lib.polyu.edu.hk/handle/200/12294}

\bibitem[{M. {Rizhko} \& J.~S. {Bloom}(2025){Rizhko} \& {Bloom}}]{2025AJ....170...28R}
{Rizhko}, M., \& {Bloom}, J.~S. 2025, \bibinfo{title}{{AstroM$^{3}$: A Self-supervised Multimodal Model for Astronomy},} \aj, 170, 28, \dodoi{10.3847/1538-3881/adcbad}

\bibitem[{A. {Roueff} {et~al.}(2024){Roueff}, {Pety}, {Gerin}, {S{\'e}gal}, {Goicoechea}, {Liszt}, {Gratier}, {Beslic}, {Einig}, {Gaudel}, {Orkisz}, {Palud}, {Santa-Maria}, {de Souza Magalhaes}, {Zakardjian}, {Bardeau}, {Bron}, {Chainais}, {Coud{\'e}}, {Demyk}, {Guzman}, {Hughes}, {Languignon}, {Levrier}, {Lis}, {Le Bourlot}, {Le Petit}, {Peretto}, {Roueff}, {Sievers}, \& {Thouvenin}}]{2024A&A...686A.255R}
{Roueff}, A., {Pety}, J., {Gerin}, M., {et~al.} 2024, \bibinfo{title}{{Bias versus variance when fitting multi-species molecular lines with a non-LTE radiative transfer model. Application to the estimation of the gas temperature and volume density},} \aap, 686, A255, \dodoi{10.1051/0004-6361/202449148}

\bibitem[{{\'A}. {S{\'a}nchez-Monge} {et~al.}(2018){S{\'a}nchez-Monge}, {Schilke}, {Ginsburg}, {Cesaroni}, \& {Schmiedeke}}]{2018A&A...609A.101S}
{S{\'a}nchez-Monge}, {\'A}., {Schilke}, P., {Ginsburg}, A., {Cesaroni}, R., \& {Schmiedeke}, A. 2018, \bibinfo{title}{{STATCONT: A statistical continuum level determination method for line-rich sources},} \aap, 609, A101, \dodoi{10.1051/0004-6361/201730425}

\bibitem[{J. Schulman {et~al.}(2017)Schulman, Wolski, Dhariwal, Radford, \& Klimov}]{schulman2017proximal}
Schulman, J., Wolski, F., Dhariwal, P., Radford, A., \& Klimov, O. 2017, \bibinfo{title}{Proximal policy optimization algorithms,} arXiv preprint arXiv:1707.06347

\bibitem[{N. {Shazeer}(2020){Shazeer}}]{2020arXiv200205202S}
{Shazeer}, N. 2020, \bibinfo{title}{{GLU Variants Improve Transformer},} arXiv e-prints, arXiv:2002.05202, \dodoi{10.48550/arXiv.2002.05202}

\bibitem[{C. Shorten \& T.~M. Khoshgoftaar(2019)Shorten \& Khoshgoftaar}]{shorten2019survey}
Shorten, C., \& Khoshgoftaar, T.~M. 2019, \bibinfo{title}{A survey on image data augmentation for deep learning,} Journal of big data, 6, 1

\bibitem[{L.~N. Smith \& N. Topin(2019)Smith \& Topin}]{smith2019super}
Smith, L.~N., \& Topin, N. 2019, in Artificial intelligence and machine learning for multi-domain operations applications, Vol. 11006, SPIE, 369--386

\bibitem[{H. {Toru Shay} {et~al.}(2025){Toru Shay}, {Scolati}, {Wenzel}, {Lee}, {Marimuthu}, \& {McGuire}}]{2025ApJ...985..123T}
{Toru Shay}, H., {Scolati}, H.~N., {Wenzel}, G., {et~al.} 2025, \bibinfo{title}{{Exploring Effects of Modified Machine Learning Pipelines of Astrochemical Inventories},} \apj, 985, 123, \dodoi{10.3847/1538-4357/adc80b}

\bibitem[{C. {Vastel} {et~al.}(2015){Vastel}, {Bottinelli}, {Caux}, {Glorian}, \& {Boiziot}}]{2015sf2a.conf..313V}
{Vastel}, C., {Bottinelli}, S., {Caux}, E., {Glorian}, J.~M., \& {Boiziot}, M. 2015, in SF2A-2015: Proceedings of the Annual meeting of the French Society of Astronomy and Astrophysics, 313--316

\bibitem[{A. Vaswani {et~al.}(2017)Vaswani, Shazeer, Parmar, Uszkoreit, Jones, Gomez, Kaiser, \& Polosukhin}]{NIPS2017_3f5ee243}
Vaswani, A., Shazeer, N., Parmar, N., {et~al.} 2017, in Advances in Neural Information Processing Systems, ed. I.~Guyon, U.~V. Luxburg, S.~Bengio, H.~Wallach, R.~Fergus, S.~Vishwanathan, \& R.~Garnett, Vol.~30 (Curran Associates, Inc.).
\newblock \url{https://proceedings.neurips.cc/paper_files/paper/2017/file/3f5ee243547dee91fbd053c1c4a845aa-Paper.pdf}

\bibitem[{T. {Villadsen} {et~al.}(2022){Villadsen}, {Ligterink}, \& {Andersen}}]{2022A&A...666A..45V}
{Villadsen}, T., {Ligterink}, N.~F.~W., \& {Andersen}, M. 2022, \bibinfo{title}{{Predicting binding energies of astrochemically relevant molecules via machine learning},} \aap, 666, A45, \dodoi{10.1051/0004-6361/202244091}

\bibitem[{P. Virtanen {et~al.}(2020)Virtanen, Gommers, Oliphant, Haberland, Reddy, Cournapeau, Burovski, Peterson, Weckesser, Bright, {van der Walt}, Brett, Wilson, Millman, Mayorov, Nelson, Jones, Kern, Larson, Carey, Polat, Feng, Moore, {VanderPlas}, Laxalde, Perktold, Cimrman, Henriksen, Quintero, Harris, Archibald, Ribeiro, Pedregosa, {van Mulbregt}, \& {SciPy 1.0 Contributors}}]{2020SciPy-NMeth}
Virtanen, P., Gommers, R., Oliphant, T.~E., {et~al.} 2020, \bibinfo{title}{{{SciPy} 1.0: Fundamental Algorithms for Scientific Computing in Python},} Nature Methods, 17, 261, \dodoi{10.1038/s41592-019-0686-2}

\bibitem[{D. Wang {et~al.}(2018)Wang, Tan, \& Liu}]{wang2018particle}
Wang, D., Tan, D., \& Liu, L. 2018, \bibinfo{title}{Particle swarm optimization algorithm: an overview,} Soft computing, 22, 387

\bibitem[{J. Wang {et~al.}(2017)Wang, Perez, {et~al.}}]{wang2017effectiveness}
Wang, J., Perez, L., {et~al.} 2017, \bibinfo{title}{The effectiveness of data augmentation in image classification using deep learning,} Convolutional Neural Networks Vis. Recognit, 11, 1

\bibitem[{ {W}es {M}c{K}inney(2010){W}es {M}c{K}inney}]{mckinney-proc-scipy-2010}
{W}es {M}c{K}inney. 2010, in {P}roceedings of the 9th {P}ython in {S}cience {C}onference, ed. {S}t\'efan van~der {W}alt \& {J}arrod {M}illman, 56 -- 61, \dodoi{10.25080/Majora-92bf1922-00a}

\bibitem[{R.~J. Williams(1992)Williams}]{williams1992simple}
Williams, R.~J. 1992, \bibinfo{title}{Simple statistical gradient-following algorithms for connectionist reinforcement learning,} Machine learning, 8, 229

\bibitem[{F. {Wyrowski} {et~al.}(2006){Wyrowski}, {Menten}, {Schilke}, {Thorwirth}, {G{\"u}sten}, \& {Bergman}}]{2006A&A...454L..91W}
{Wyrowski}, F., {Menten}, K.~M., {Schilke}, P., {et~al.} 2006, \bibinfo{title}{{Revealing the environs of the remarkable southern hot core G327.3-0.6},} \aap, 454, L91, \dodoi{10.1051/0004-6361:20065347}

\bibitem[{Z. Zhang(2023)Zhang}]{Zhang_2023}
Zhang, Z. 2023, \bibinfo{title}{{PRIMA: Reference Implementation for Powell's Methods with Modernization and Amelioration},}, available at http://www.libprima.net, DOI: 10.5281/zenodo.8052654

\end{thebibliography}
\bibliographystyle{aasjournalv7}



\end{document}